\documentclass[sigconf,nonacm]{acmart} 
\acmSubmissionID{1006}

\usepackage{booktabs} 
\usepackage[export]{adjustbox} 
\usepackage{tikz}
\ifcsname intercal\endcsname
\renewcommand{\intercal}{T}
\else
\newcommand{\intercal}{T}
\fi

\usetikzlibrary{positioning,calc,shadows,shadows.blur}
\usepackage{pgfplots}
\pgfplotsset{compat=1.6}

\usepackage{color, colortbl}
\definecolor{myBlue}  {rgb}{0.302,0.345,0.631}
\definecolor{myGreen} {rgb}{0.125,0.651,0.329}
\definecolor{myRed}   {rgb}{0.651,0.125,0.329}

\newcommand{\update}[1] {{#1}}

\usepackage[ruled]{algorithm2e} 

\SetAlFnt{\small}
\SetAlCapFnt{\small}
\SetAlCapNameFnt{\small}
\SetAlCapHSkip{0pt}

\begin{document}
\title{A Data-Driven Paradigm
      for Precomputed Radiance Transfer}

\author{Laurent Belcour}
\orcid{0000-0001-5208-5622}
\affiliation{%
 \institution{Unity Technologies}
}

\author{Thomas Deliot}
\orcid{0000-0001-5208-5622}
\affiliation{%
 \institution{Unity Technologies}
}

\author{Wilhem Barbier}
\orcid{0000-0001-5208-5622}
\affiliation{%
 \institution{ENSIMAG}
}

\author{Cyril Soler}
\affiliation{%
 \institution{Inria}
}
\settopmatter{authorsperrow=4}
\renewcommand\shortauthors{Belcour L. et al}

\begin{abstract}
    In this work, we explore a change of paradigm to build \textit{Precomputed Radiance Transfer} (PRT) methods in a data-driven way. This paradigm shift allows us to alleviate the difficulties of building traditional PRT methods such as defining a reconstruction basis, coding a dedicated path tracer to compute a transfer function, etc. Our objective is to pave the way for Machine Learned methods by providing a simple baseline algorithm. More specifically, we demonstrate real-time rendering of indirect illumination in hair and surfaces from a few measurements of direct lighting. We build our baseline from pairs of direct and indirect illumination renderings using only standard tools such as Singular Value Decomposition (SVD) to extract both the reconstruction basis and transfer function.
\end{abstract}

%
%
    

%
%


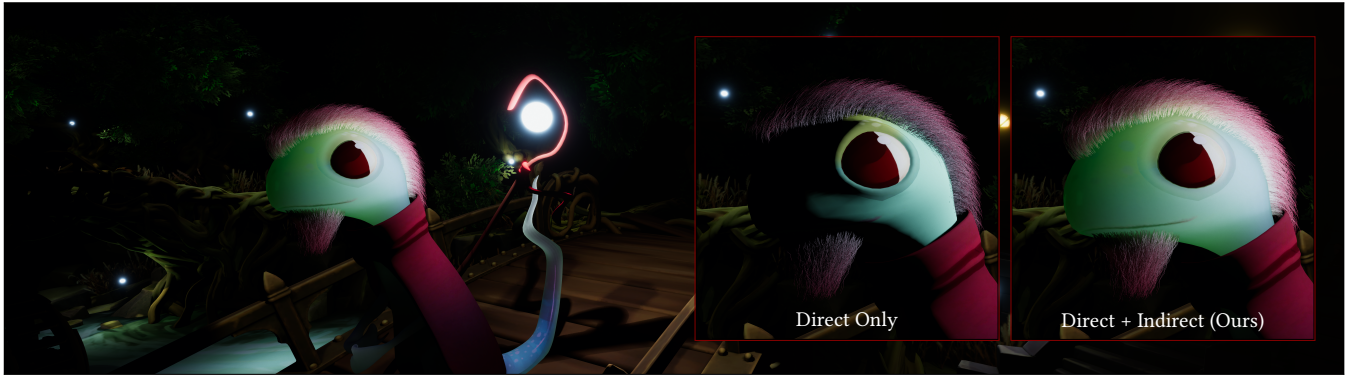
\begin{teaserfigure}
    \centering
    \begin{tikzpicture}[font=\small]
        \node[inner sep=0pt] { \includegraphics[width=\linewidth,frame,clip,trim=0 32 0 0]{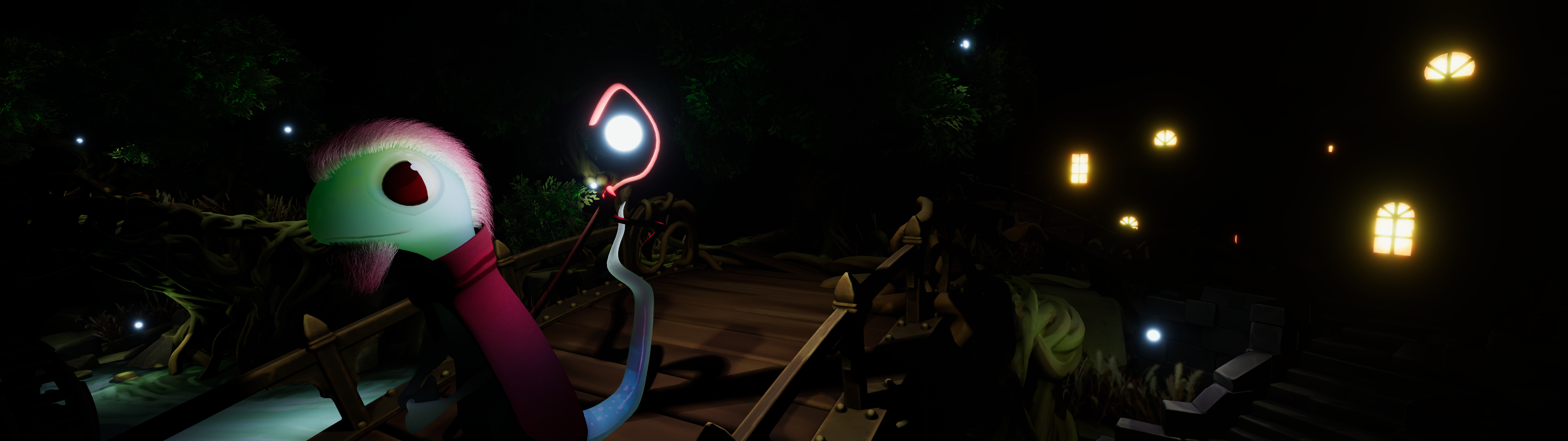} };
        %
        \node[inner sep=0pt] (A) at (6.5cm,0cm){ \includegraphics[height=4cm,frame]{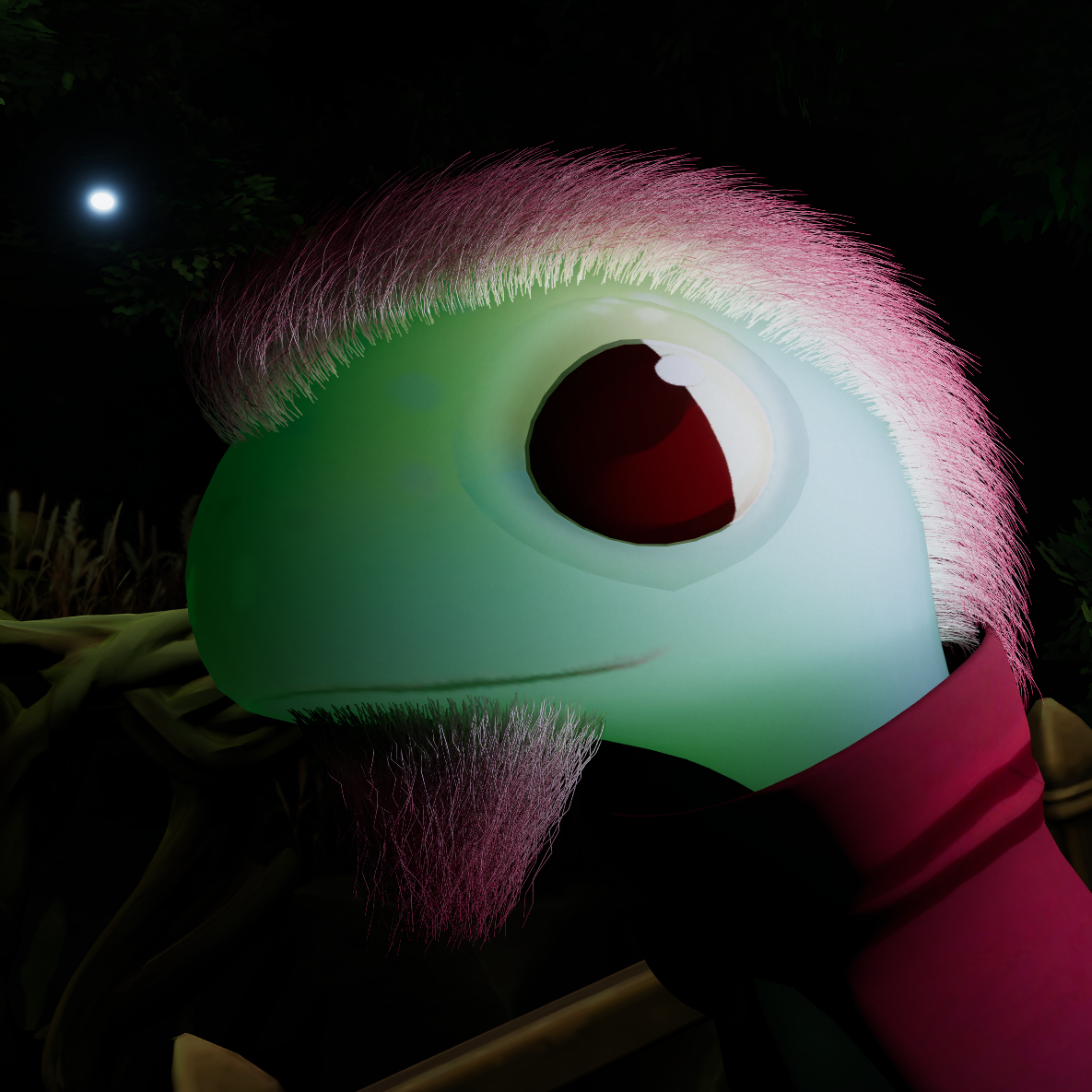} };
        \draw[color=red!50!black] (A.south west) rectangle (A.north east);
        \node[below=-14pt of A.south] { \textcolor{white}{Direct + Indirect (Ours)} };
        Inset B (ours)
        \node[inner sep=0pt] (B) at (2.3cm,0cm){ \includegraphics[height=4cm,frame]{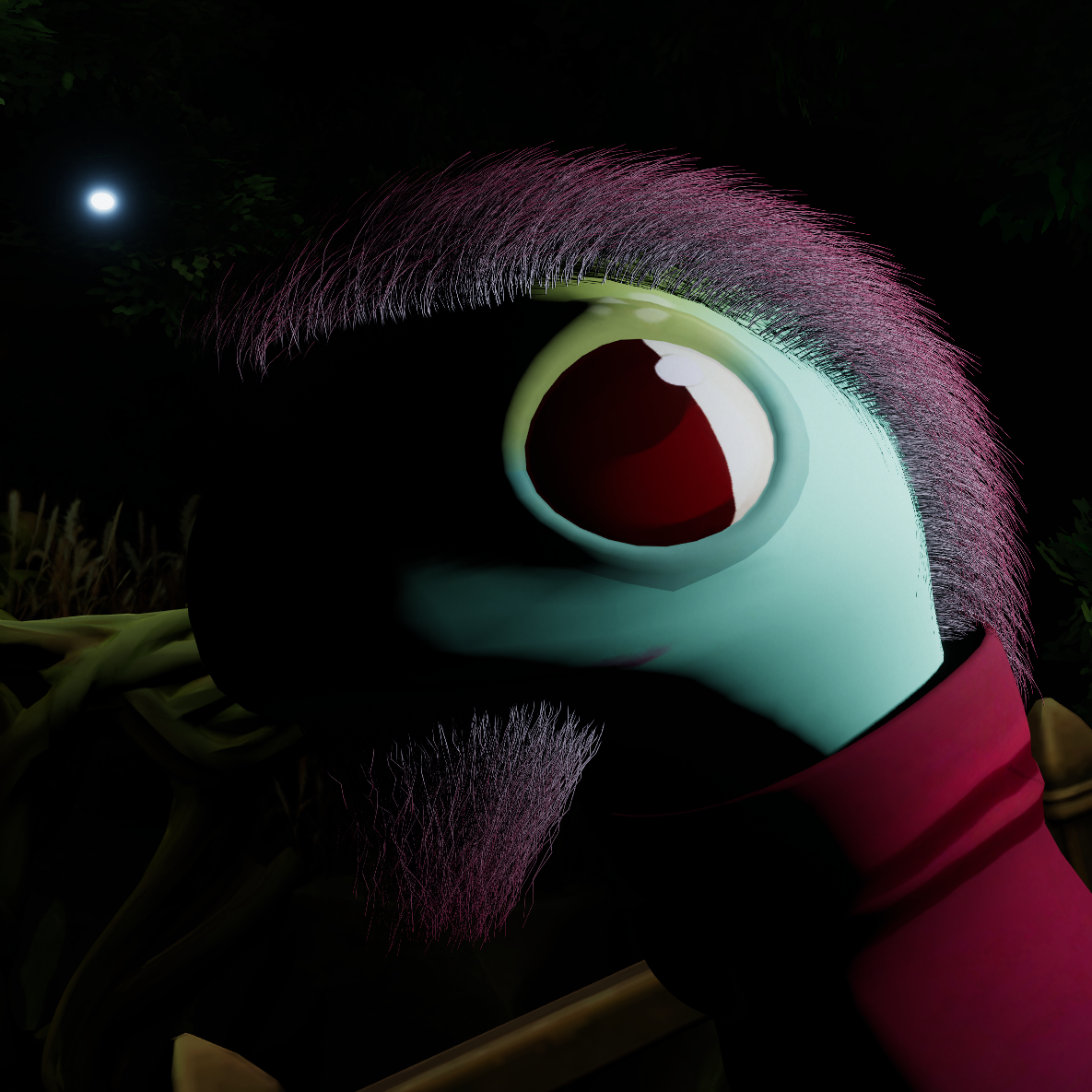} };
        \draw[color=red!50!black] (B.south west) rectangle (B.north east);
        \node[below=-14pt of B.south] { \textcolor{white}{Direct Only} };
    \end{tikzpicture}
    \vspace{-15pt}
    \caption{\textbf{Real-Time Direct to Indirect Transfer}. \textnormal{We tackle the problem of rendering in real-time effects such as indirect illumination in hair, surfaces, or volumes on key assets such as this Lizard mage. We build a data-driven formulation of Precomputed Radiance Transfer and construct a practical baseline algorithm using this paradigm. We further demonstrate its use in a commercial game engine.}
    \label{fig:teaser}
    }
\end{teaserfigure}

\maketitle

%
%
\section{Introduction}

\paragraph{Motivation}
Despite the recent introduction of hardware ray-tracing, full real-time global illumination (GI) remains too expensive to solve for modern video-game engines. As a result, many local GI effects such as subsurface scattering or indirect illumination in hair are still handled using alternative techniques that are specific to a particular effect~\cite{Golubev2018,Tafuri2019}. In this work, we are interested in generic solutions for real-time low-frequency GI that do not require ray tracing capabilities at runtime. \\

\textit{Precomputed Radiance Transfer} (PRT) is the ideal candidate to evaluate indirect illumination. It is both efficient (runs in real-time) and generic (handles different kinds of transport).
PRT methods~\cite{sloan2002precomputed} store a light transport matrix at each vertex of the geometry in order to reproduce, at runtime, indirect illumination given the incident illumination. During rendering, this matrix translates the incident illumination into outgoing indirect illumination.  While those methods solve light transport effectively, they come at a cost in their design. They require: a user defined (basis) representation for radiance; a tedious and dedicated precomputation stage; an advanced clustering method to cope with memory cost, \textit{etc}.
Therefore, we investigated whether a solution free of those constraints was achievable.

\paragraph{Data-Driven Approaches}
It would be temping to alleviate those constraints using Machine Learning (ML) approaches. Indeed, using a dataset, an optimization process could construct a transfer function and an internal representation for direct to indirect transport. However, the trade-off between complexity and output quality would be hard to assess. To pave the way for such approaches, we need a simple baseline to assess the efficiency of ML methods when benchmarking them. Our aim is to provide such baseline.

\paragraph{A New Paradigm for PRT}
In this paper, we show that a different paradigm for PRT liberates us from computing the transfer matrix as well as the radiance representation. Inspired by ML's formalism, we express PRT in a data-driven paradigm. While we retain the advantages of ML approaches, it allows us to construct a lean baseline method. Such baseline defines a reference point that more complex ML-based approaches can benchmark against.

\paragraph{Outline}
We structured our contributions as follows:
\begin{itemize}
    \item In Section~\ref{sec_ddrt}, we introduce a data-driven formulation of direct-to-indirect transfer and show that classical PRT can be inferred solely from data.
    \item In Section~\ref{sec_baseline}, we introduce a simple algorithm: our proposed baseline. Specifically, we describe a meshless radiance transfer algorithm that takes only a few measurements of direct illumination to infer indirect illumination in texture space.
    \item In Section~\ref{sec_details}, we describe practical details to avoid temporal aliasing, support high dynamic of light and animation. 
    \item In Section~\ref{sec_results}, we validate that our method runs in real-time, and support multiple transport scenario (surfaces, volumes, hair).
\end{itemize}

%
%
\section{Previous Work}

In this section, we review some of the most relevant works for solving GI effects in real time. 
For a more thorough survey of real-time GI, we refer the reader to the survey of 
Ritschel~et~al.~\shortcite{ritschel2012state}.
Modern game engines usually implement various methods for computing and/or approximating 
global illumination under the performance constraints of real-time rendering.

\paragraph{Static Global Illumination} Lightmaps store the result of diffuse global illumination computed offline~\cite{abrash2000quake,Silvennoinen2019,seyb20uberbake}. While very efficient, this approach imposes static lighting at runtime and cannot reproduce sophisticated optical effects such as subsurface scattering. \\

\textit{Path-Tracing} provides a means to solve global illumination using Monte Carlo integration~\cite{kajiya1986rendering}. However, small effects such as indirect illumination in hair or highly diffusing participating media can take little space on screen, but require significant computing efforts to converge. While path tracing can be accelerated using techniques such as using virtual point lights~\cite{keller1997instant}, importance sampling~\cite{bitterli2020spatiotemporal}, radiance caching~\cite{muller2021real} or denoising~\cite{icsik2021interactive}, those methods often target global variance reduction and would struggle with localized indirect light transport. More dedicated methods exist to approximate light transport, but often requires a fair amount of runtime power.

\paragraph{Precomputed Radiance Transfer}
PRT~\cite{sloan2002precomputed,sloan2003clustered,sloan2005local} computes per-vertex transfer operators that map incident radiance projected in a basis to outgoing radiance. Such methods can simulate both direct (soft shadows) and indirect effects (indirect illumination) on static objects from any distant light illumination settings. However, such generality comes at a cost. For example, one must choose a basis representation for incident lighting. Many have been tested such as Spherical Harmonics~\cite{sloan2002precomputed,ramamoorthi2001efficient}, Spherical Gaussians~\cite{green2006view,tsai2006all}, and Wavelets~\cite{ng2003all}.

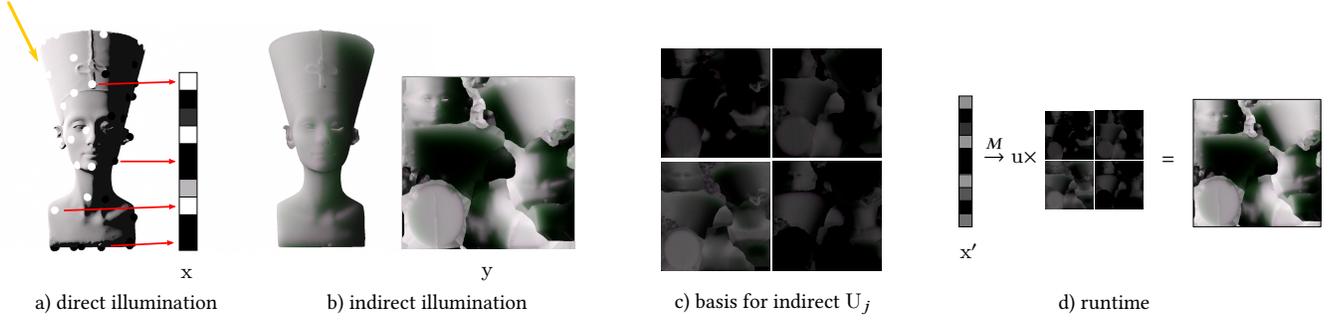
\begin{figure*}[t!]
    \begin{tikzpicture}[font=\small]
        \node {
            \adjincludegraphics[width=\linewidth]{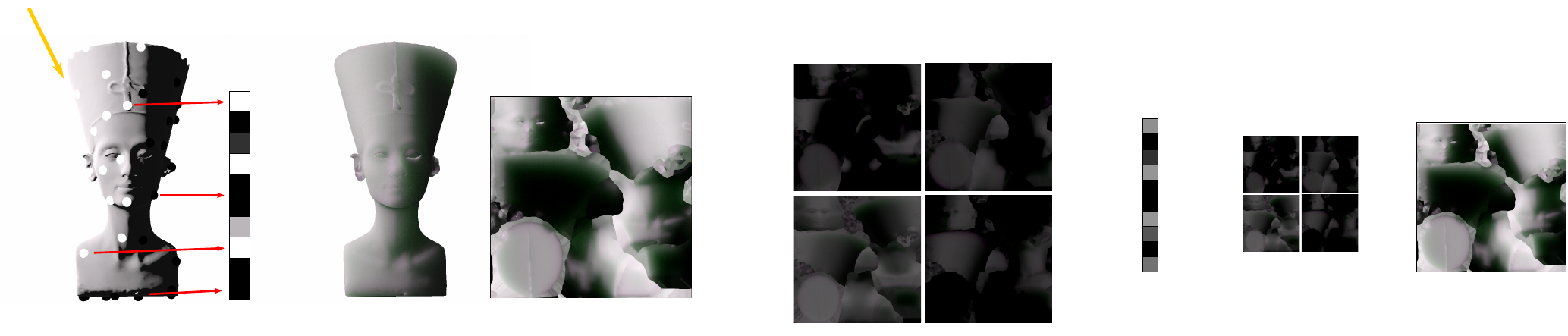}
        };
        \node at (-7.0, -2.2) {a) direct illumination};
        \node at (-6.2, -1.8) {$ \mathbf{x} $};
        \node at (-3.0, -2.2) {b) indirect illumination};
        \node at (-2.2, -1.8) {$ \mathbf{y} $};
        \node at (+1.6, -2.2) {c) basis for indirect $\mathbf{U}_j$};

        \node at (+4.20, -1.5) { $ \mathbf{x}^\prime $};  
        \node at (+4.75, -0.15) { $\overset{M}{\rightarrow} \mathbf{u} \times$};
        \node at (+6.85, -0.3) { $=$}; 
        \node at (+6.0, -2.2) {d) runtime};
    \end{tikzpicture}
    \vspace{-20pt}
    \caption{
        Outline of Data-Driven PRT.
        \textnormal{
        A lighting condition gives us a vector of direct illumination sampled around the object $\mathbf{x}$ (a), and a baked indirect illumination in texture space $\mathbf{y}$ (b). Randomizing the lighting condition, we extract a set of basis elements for the indirect illumination (c). From it, we compute a transfer matrix $M$ that enables to compute coefficients $\mathbf{u} = M \times \mathbf{x}$ of indirect illumination (d).
        }
        \label{fig:pipeline}
    }
\end{figure*}

\paragraph{Geometric Constraints} 
Typically, PRT works by linearly interpolating per vertex outgoing radiance.
This naturally makes its storage and computational cost proportional to the amount of geometric details.
A hierarchical basis on meshes~\cite{lehtinen2008meshless} or in screen-space~\cite{havsan2006direct} can avoid this problem.
However, those bases must resolve all possible illuminations. It makes them over-generic when only a subset of lighting conditions are experienced at runtime. Furthermore, they require the explicit computation of the transfer matrix.
In our work, both the basis and the transfer matrix are extracted from the dataset. Furthermore, our meshless formulation is compatible with both volumetric, subsurface, or complex geometries without a theoretical change.

\paragraph{PRT Optimized for Lighting Scenarios} Closest to our work are Modular Radiance Transfer (MRT)~\cite{loos2011modular} and Reduced Aggregate Scattering Operators (RASO)~\cite{blumer2016reduced}. Both used a custom lighting scenario to obtain better basis representation for direct illumination. However, in a second step they estimate the indirect illumination basis and the transfer function using the classical approach. We show that such a two-steps approach is not necessary and embrace a fully data-driven approach.

%
%
\section{Data-Driven Radiance Transfer}
\label{sec_ddrt}

Given a dataset of $N$ direct illuminations $\left\{ \mathbf D_k \right\}_{0 \dots N}$ (see Figure~\ref{fig:pipeline}~(a)) along with the corresponding full indirect illuminations $\left\{ \mathbf I_k \right\}_{0 \dots N}$ (see Figure~\ref{fig:pipeline}~(b)) we want to express the direct-to-indirect transfer:
\begin{align}
    \mathbf{I}_k = f(\mathbf{D}_k) \, \forall k \in [0, N].
\end{align}
We do not yet explicit the nature of direct and indirect illumination data, but we treat them as vectors. These can be either volumetric, point-based measurements, textures, \textit{etc}. In the following section, we describe how to build the necessary elements for Precomputed Radiance Transfer from this dataset.

\subsection{Estimating the Transfer Matrix}

We denote as $D$ (resp. $I$) the matrix obtained by packing the direct (resp. indirect) illumination data $\mathbf D_k$ (resp. $\mathbf I_k$) as columns.
Since light transport is a linear operator, there exists a linear mapping between the direct and indirect illuminations in a scene. Thus, there is a light transport matrix $M$ for which
\begin{align}
    I = M D.
\end{align}
Depending on how direct and indirect illuminations are encoded, matrices $I$ and $D$ can be very large matrices, and $M$ is hence a potentially very large matrix. We assume that the dataset of direct lighting can compactly represent any lighting configuration. Hence, any given direct illumination vector $\mathbf x$ can be expressed as a linear combination of the columns of $D$ with coefficients $\mathbf c\in\mathbb{R}^N$, we have
\begin{align}
    \mathbf{x} = D \mathbf{c},
    \label{eq:yDx}
\end{align}
and by multiplying by the transfer matrix $M$, we obtain:
\begin{align}
    M\mathbf x = M D \mathbf c = I\mathbf c.
    \label{eq:My_eq_Ix}
\end{align}
Hence, the indirect illumination caused by $\mathbf x$ can be computed as a linear combination of the columns of $I$.
In other words, computing $M$ is not necessary but it requires us to know the value of $\mathbf c$. We can avoid its direct calculation using Equation~\ref{eq:yDx} and expressing $\mathbf{c}$ with respect to $\mathbf{x}$:
\begin{align}
    \mathbf{c} = (D^\intercal D)^{-1} D^\intercal \mathbf{x}.
\end{align}
Injecting this formula into Equation~\ref{eq:My_eq_Ix}, we obtain the formula of the indirect illumination $\mathbf{y}$ w.r.t. direct illumination $\mathbf{x}$:
\begin{align}
	\mathbf{y} = M\mathbf{x} = I (D^\intercal D)^{-1} D^\intercal \mathbf{x}.
	\label{eqn:My}
\end{align}
This expresses the light transport matrix as an ordinary least squares problem. It can also be seen as the transfer of the projection of $\mathbf{x}$ in the dataset of direct illumination: $\mathbf{x}^{\perp} = D^{T} \mathbf{x}$.

\begin{figure}[b]
    \begin{tikzpicture}[font=\footnotesize]
        \begin{semilogyaxis}[
            grid=major,
            width = \linewidth,
            height = 4cm,
            xmin = 0.0, xmax = 32.0,
            ymin = 0.0, ymax = 1.0,
            xlabel = {number of eigenvectors},
            ylabel = {relative eigenvalue},
            xtick = {0,4,8,16,32,64,128},
            legend pos = south east,
            legend style= {
                    at={(1.05, 0.6)},
                    nodes= {
                        scale=0.8,
                        transform shape
                    }
                },
            ]
            \addplot[black!00!cyan,line width=1.2pt] table {figures/eigenvalues/nefertiti_eig.txt};
            \addlegendentry{\textsc{Nefertiti}}
            \addplot[black!00!orange,line width=1.2pt] table {figures/eigenvalues/viking_eig.txt};
            \addlegendentry{\textsc{Viking}}
            \addplot[black!20!green,line width=1.2pt] table {figures/eigenvalues/hairball_eig.txt};
            \addlegendentry{\textsc{Hairball}}
            \addplot[red!20!pink,line width=1.2pt] table {figures/eigenvalues/dragon_eig.txt};
            \addlegendentry{\textsc{Dragon}}
        \end{semilogyaxis}
    \end{tikzpicture}
    \vspace{-10pt}
    \caption{
        \label{fig:eigenvalues}
        Relative Eigenvalues of the indirect illumination.
        \textnormal{We display the relative eigenvalues of the aucorrelation matrix of the dataset of indirect illumination vectors for the different test scenes. We found out that they drop quickly below $1\%$ and only a few dimensions are relevant for reconstruction.}
    }
\end{figure}
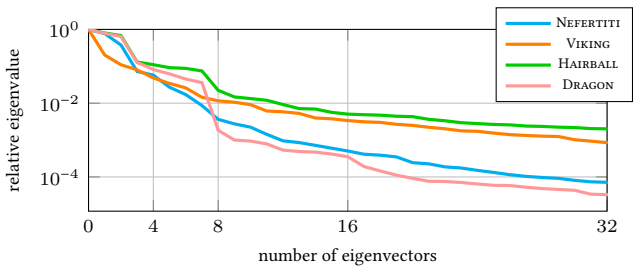

\subsection{Indirect Illumination Basis Extraction}
In Equation~\ref{eqn:My}, the columns in $I$ are used as a non-orthogonal basis to approximately represent the space spanned by $M$. This basis may have too many elements for practical use. However, similar to \citet{loos2011modular} we found that the space of indirect illumination is effectively low dimensional (see Figure~\ref{fig:eigenvalues}). We solve this problem using the singular value decomposition of $I$:
\begin{align} I = U \Sigma V^\intercal.
	\label{eqn:svd}
\end{align}
In this decomposition, the columns of $U$ form an orthogonal basis (see Figure~\ref{fig:pipeline}~(c)), and the diagonal elements in $\Sigma$ measure the importance of each of these vectors in representing this space. We approximate $I$ by keeping only the $n$ most important diagonal elements of $\Sigma$, and write our approximation of Equation~\ref{eqn:svd} as
\begin{align}
    I \approx U_n \, \Sigma_n \, V_n^\intercal = U_n \, C_n,
\end{align}
with $C_n$ the matrix of coefficients. Plugging in Equation~\ref{eqn:My}, we get
\begin{align}
	\mathbf{y} &\simeq U_n C_n (D^\intercal D)^{-1} D^\intercal \mathbf{x} \\
    &\simeq U_n {M}_n \mathbf{x},
	\label{eqn:My2}
\end{align}
with ${M}_n = C_n (D^\intercal D)^{-1} D^\intercal$. \\

The above equation computes the coefficients $\mathbf{u} = {M}_n \mathbf{x}$ of indirect illumination using a set of orthogonal basis vectors (columns of $U_n$) that are tailored to the lighting conditions seen in the matrix $D$ and $I$. So far we have reduced the output of the transport matrix using the SVD. To make our method fully practical, we need to ensure that the dimension of the direct vectors is constrained.

\subsection{Direct Vectors}
\label{sec:direct_vectors}
Traditional PRT methods do not compute the transfer matrix between direct and indirect illumination. They rather store the transfer between incident illumination and indirect illumination. Incident illumination is expressed using a finite number of Spherical Harmonics coefficients $\mathbf{w}$. In such a case, we can write the transport operator that maps SH coefficients into direct illumination:
\begin{align}
    \mathbf{x} = T \mathbf{w}.
\end{align}
Thus, we can write
\begin{align}
    \mathbf{y} = M T \mathbf{w}.
\end{align}
As long as $T$ defines a bijection between $\mathbf{x}$ and $\mathbf{w}$ (that is, it is invertible), the transfer matrices $M$ and $M \times T$ are equivalent. Thus, we have described the same transfer function as Precomputed Radiance Transfer for indirect illumination. It follows that the placement of direct illumination is critical in both the conditioning of the transfer matrix, and the achievable equivalent incident illumination.

\subsection{Discussion w.r.t. PRT}

\paragraph{Meshless Transport}
Our formalism for data-driven PRT is not tied to the geometry of the asset. While we could follow traditional PRT methods and compute transfer at vertices of the mesh, our formalism permits decoupling them. In practice, we advise for a meshless approach as reconstructed high frequencies are not linked to geometrical density.

\paragraph{Baking the Transport Matrix} PRT methods bake the transport matrix using implicit light sources defined by the illumination basis. Those light sources shade the asset with positive and negative radiance values. Hence, a dedicated light transport algorithm is used for them. In contrast, our formulation does not need to evaluate such lights and can take advantage of any advances in rendering algorithms (importance sampling, guiding, ...).

\paragraph{Lighting Agnostic Basis} While traditional PRT methods approximate the light transport matrix $M$ in a basis made of data-agnostic elements (e.g. basis functions such as wavelets, polynomials, etc.), our method designs its own basis elements (the columns of $U_n$) to represent the output of $M$ without explicitly encoding $M$.  In essence, this is similar to random low-rank matrix approximation~\cite{halko2011finding}: we build a space representing the output of $M$ from \textit{random} inputs to the light transport operator.
This data-dependent approach also brings us a major advantage over light-agnostic PRT: our representation of the illumination is not limited in frequency in regions where the indirect illumination 
changes abruptly and does not waste discretization elements in regions where the illumination is always smooth.

\begin{figure}[b]
    \centering
	\vspace{-10pt}
    \begin{tikzpicture}[font=\small]
        \node[]               (C) { \includegraphics[height=4cm]{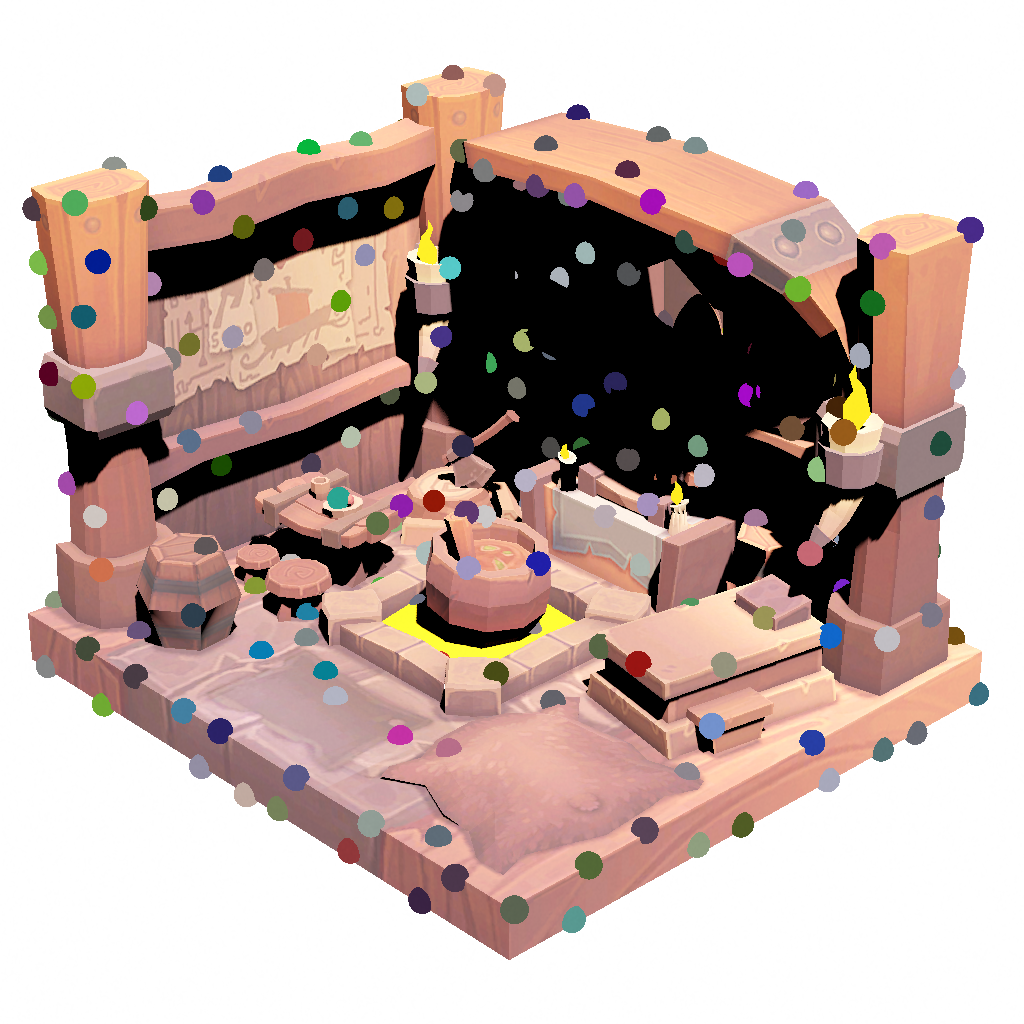} };
        \node[right=0pt of C] (D) { \includegraphics[height=4cm]{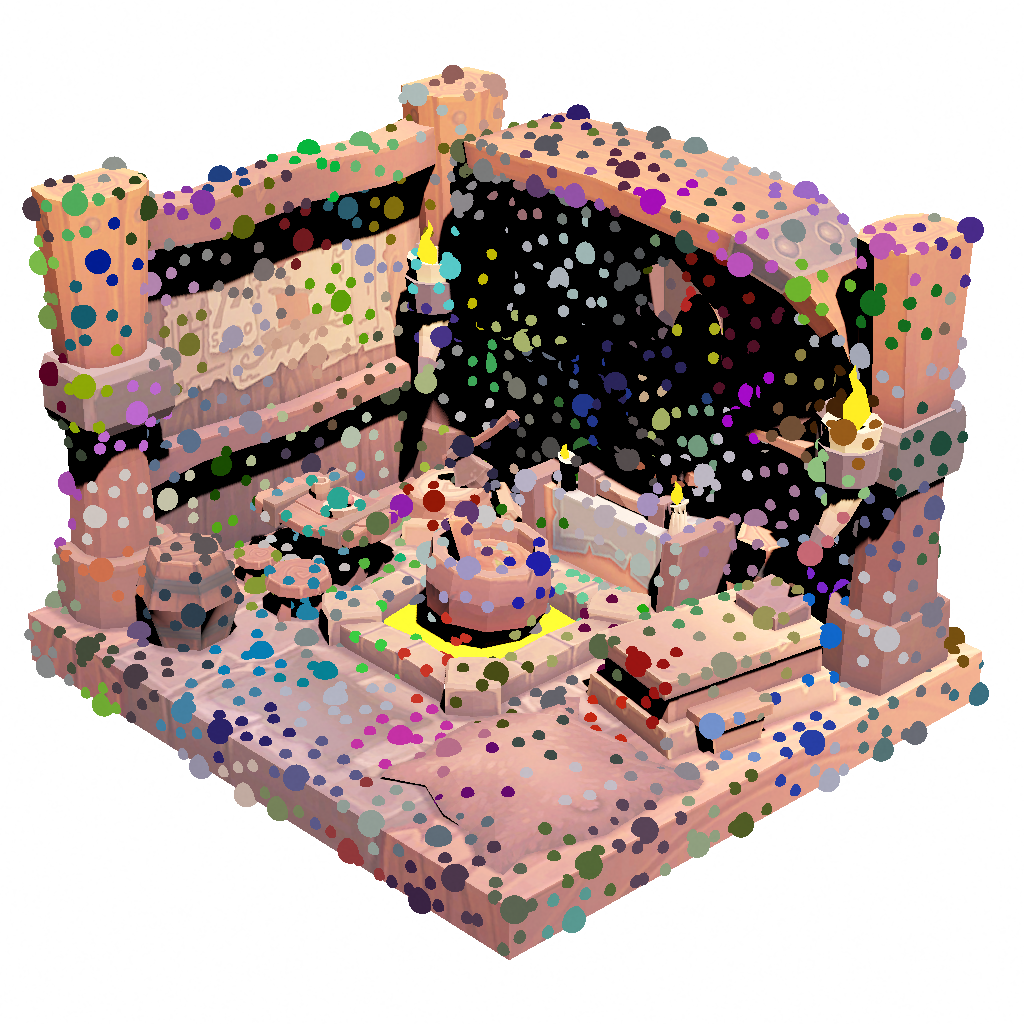} };

        \node[below=-5pt of C] {a) Measurement points};
        \node[below=-5pt of D] {b) $8\times$ sampling};
    \end{tikzpicture}
	\vspace{-10pt}
    \caption{
        Measurement points.
        \textnormal{
        We measure direct illumination at punctual positions in the scene. To better condition the incident irradiance vector, we distribute the measurement points as a blue noise on the target shape (a). To reduce flickering during runtime, we average the contribution of neighboring points (b). We show a super-sampling of $8 \times$ with averaging color coded.
        }
        \label{fig:measurement}
    }
\end{figure}

\section{A Lean Baseline}
\label{sec_baseline}

In this section, we describe the baseline method we designed. First, we detail our baking process. It follows two steps: first the database creation (Section~\ref{sec_database}), then the basis and transfer matrix extraction (Section~\ref{sec_extraction}). Then, we detail the runtime component (Section~\ref{sec_rendering}).

\subsection{Database Creation}
\label{sec_database}
In this first stage, we sample many random lighting configurations and render both direct and indirect illumination to build matrices $D$ and $I$. This step setups lights positions, orientations and intensities around the asset for each configuration. Light interaction with the asset is evaluated using a path tracer for indirect transport and the runtime renderer for direct illumination. We evaluate indirect illumination in textures using the UV-space of the asset, thus each row of $I$ is a lightmap. We evaluate direct illumination at discrete points we call \textit{measurement points}. Each measurement point consists of a position, normal, and material information (for example, albedo). Depending on the type of geometry considered, we place them either on the surface of the object (for subsurface scattering) or on the geometry hull (for hair). Other patterns could be used as well, but we did not experiment with them. We store the measurement points for later use during runtime (see Section~\ref{sec:runtime_gbuffer}).

\paragraph{Measurement Points Distribution} We distribute measurement points evenly on the target surface (either the object or its convex hull). This distribution better conditions the Gram matrix $D D^T$. Measurement points too close to each other could produce the same lighting contribution and prevent us from inverting this matrix. To produce a blue-noise distribution of measurement points we use sample elimination~\cite{yuksel2015sample}. This method selects a subset of a denser set of uniformly distributed random points to produce an even distribution as shown in Figure~\ref{fig:measurement}~(a).

\subsection{Basis and Transfer Matrix Extraction}
\label{sec_extraction}
Once the database is computed, we first extract the basis $U_n$ using Singular Value Decomposition (SVD) and then compute the direct to indirect transfer matrix $M_n$.

\paragraph{Indirect Illumination Basis}
In our experiments, we always had more pixels than examples in matrix $I$. Thus, we extract $U$ from the covariance $I^T \times I$, a $N\times N$ matrix~\cite{turk1991eigenfaces}. To perform the SVD, we always subtract the mean of the indirect illumination data as this is common practice. Since the basis works with mean subtracted images we need to add it back afterwards, at runtime. So we store the mean along with the basis. Hence, when we report using $16$ basis elements, we effectively use the mean and $15$ basis elements. To work with any light intensity, we always normalize the indirect vector by the norm of the direct vector. At runtime, we scale the mean with the norm of the direct vector.

\paragraph{Transfer Matrix} Once the basis is computed, we use Equation~\ref{eqn:My2} to compute ${M}_n$. Despite the measurement points distribution, the inverse covariance matrix of the direct illumination $\left[ D D^T \right]^{-1}$ can be ill-posed. To overcome this, we always use the Moore-Penrose pseudo inverse~\cite{penrose1955generalized} to perform it:
\begin{align}
    M_n \simeq C \left[ D D^T  \right]^{+} D^T
\end{align}

\subsection{Runtime Rendering}
\label{sec_rendering}

Our runtime algorithm works solely on GPU: first, we render the direct illumination vector; then, in a compute shader, we evaluate the coefficients $\mathbf{u}$ for the indirect illumination; finally, in the fragment shader, we blend the basis elements with these coefficients.

\paragraph{Fake G-Buffer Rendering} \label{sec:runtime_gbuffer} We use the set of measurement points as a G-Buffer that we shade using Deferred Shading~\cite{saito1990comprehensible}. This works with any lighting type (point lights, area lights, environments lights, ...) and easily integrates into modern rendering engines.

\paragraph{Deringing} Because we reconstruct indirect illumination with a few basis elements, ringing artifacts (oscillation with negative values) may occur. Since our formulation expresses the transfer matrix as an ordinary least-squares problem, we add a Tikhonov regularization (or ridge regression) term to Equation~\ref{eqn:My}:
\begin{align}
    M_n \simeq C \left[ D D^T + \Lambda \right]^{-1} D^T
\end{align}
where $\Lambda$ is a diagonal matrix. We usually use $\Lambda = \lambda \, I$, where $I$ is the identity matrix and $\lambda$ is a small scalar.

%
%
\section{Practical Details}
\label{sec_details}

\subsection{Direct Illumination Downscaling}
The length of the direct illumination vector defines a trade-off between performance and quality.
Using only a few measurement points enables us to efficiently evaluate the direct illumination using Deferred Shading and perform the matrix-vector product ${M}_n  \mathbf{x}$. However, too few measurement points will result in aliasing in both the direct illumination evaluation and the indirect reconstruction. We choose to counter aliasing by increasing the number of measurement points when doing shading (see Figure~\ref{fig:measurement}~(b)), but downscaling the vector of direct illumination afterwards. We do so by averaging neighboring direct samples when the measurement points are geometrically close (similar to mipmapping). We opted for a compute-shader pass that downsamples an arbitrary number of measurement points per pixel for this task.

\begin{figure}[t]
    \centering
    \begin{tikzpicture}[font=\small]
        \begin{scope}
            \draw (0.25,0) rectangle (0.45,2);
            \node at (0.75,1) { $\times$ };
            \draw (1,0) rectangle node { $T_{RGB}$ } (3,2);
            \node at (3.25,1) { $=$ };
            \draw (3.55,0) rectangle (3.75,2);

            \draw[fill=blue]  (0.25,8/10)  rectangle (0.45,10/10);
            \draw[fill=green] (0.25,10/10) rectangle (0.45,12/10);
            \draw[fill=red]   (0.25,12/10) rectangle (0.45,14/10);
            \draw[fill=blue]  (0.25,14/10) rectangle (0.45,16/10);
            \draw[fill=green] (0.25,16/10) rectangle (0.45,18/10);
            \draw[fill=red]   (0.25,18/10) rectangle (0.45,20/10);
            \node at (0.35,0.5) { $\vdots$ };

            \draw[fill=blue]  (3.55,8/10)  rectangle (3.75,10/10);
            \draw[fill=green] (3.55,10/10) rectangle (3.75,12/10);
            \draw[fill=red]   (3.55,12/10) rectangle (3.75,14/10);
            \draw[fill=blue]  (3.55,14/10) rectangle (3.75,16/10);
            \draw[fill=green] (3.55,16/10) rectangle (3.75,18/10);
            \draw[fill=red]   (3.55,18/10) rectangle (3.75,20/10);
            \node at (3.65,0.5) { $\vdots$ };
        \end{scope}
        \node at (2,-0.3) { a) multi-channel basis };

        \begin{scope}[xshift=4.5cm]
            \draw[fill=blue]  (0.25,0)   rectangle (0.45,2/3);
            \draw[fill=green] (0.25,2/3) rectangle (0.45,4/3);
            \draw[fill=red]   (0.25,4/3) rectangle (0.45,6/3);

            \node at (0.75,1) { $\times$ };

            \draw (1,0) rectangle (3,2);
            \draw[fill=red]   (1+0/3,4/3) rectangle node { $T_{R}$ } (1+2/3,6/3);
            \draw[fill=green] (1+2/3,2/3) rectangle node { $T_{G}$ } (1+4/3,4/3);
            \draw[fill=blue]  (1+4/3,0/3) rectangle node { $T_{B}$ } (1+6/3,2/3);

            \node at (3.25,1) { $=$ };

            \draw[fill=blue]  (3.55,0/3) rectangle (3.75,2/3);
            \draw[fill=green] (3.55,2/3) rectangle (3.75,4/3);
            \draw[fill=red]   (3.55,4/3) rectangle (3.75,6/3);
        \end{scope}
        \node at (6.5,-0.3) { b) mono-channel basis };
    \end{tikzpicture}
    \caption{
        \textbf{Multi vs Mono channel basis.}
        \textnormal{
            In our framework, we can either encode in the basis the color channels or separate them. In the latter case, this results in applying three different transfer matrices to the incident illumination by separating the color channels.
        }
        \label{fig:multichannel-basis}
    }
\end{figure}
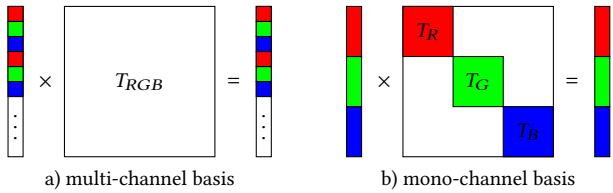

\subsection{Mono or Multi-Channel Basis}
So far, we assumed that the direct and indirect illumination, $\mathbf{D}_k$ and $\mathbf{I}_k$, are vectors. They do include color information using an interleaved pattern: $\mathbf{I}_k = \left[ \mathbf{I}^R_{k,0}, \mathbf{I}^G_{k,0}, \mathbf{I}^B_{k,0}, \dots \right]$. Hence, the basis decomposition of the indirect lighting data will incorporate color reconstruction and by construction the transfer matrix as well. This can be detrimental when trying to change the color of the light sources with colors not seen in the dataset. In practice, we treat color channels as separate vectors to build a unique reconstruction basis for all color channels. However, using such splitting, we have to perform a transfer matrix multiplication per color channel (see Figure~\ref{fig:multichannel-basis}).

\paragraph{GPU Storage and Evaluation} Since each basis element is used to reconstruct the three color channels, we pack them into RGBA textures to store four basis elements per texture. At runtime, we upload the reconstruction coefficient as $4 \times 3$ matrices on the GPU and perform the matrix/vector multiplication in the fragment shader to get the output color.

\subsection{Working in Gamma Space}
\begin{figure}[t]
    \begin{tikzpicture}[font=\small]
        \node { \includegraphics[width=\linewidth,frame]{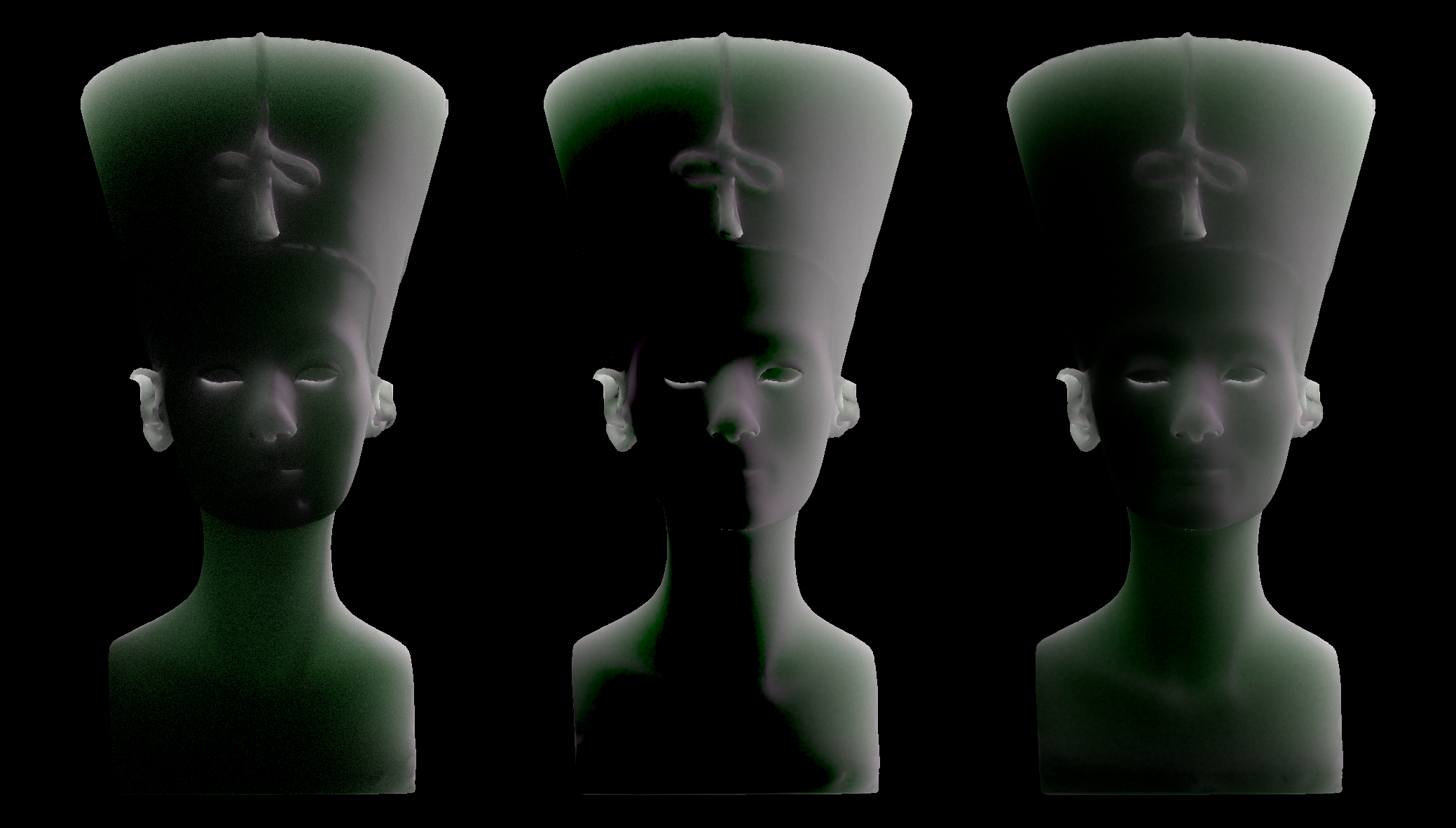} };
        \node at (-2.7, -2) { \textcolor{white}{Reference} };
        \node at ( 0.0, -2) { \textcolor{white}{Ours - linear} };
        \node at ( 2.7, -2) { \textcolor{white}{Ours - gamma} };
    \end{tikzpicture}
    \vspace{-20pt}
    \caption{
        \textbf{Working in Gamma space.}
        \textnormal{
            We perform transport and reconstruction in Gamma space to overcome the inability of the basis decomposition to cope with high contrast such as this side light. We found that this resolves hard discontinuity as shown in the middle rendering.
        }
        \label{results:nefertiti}
        \vspace{-10pt}
    }
\end{figure}
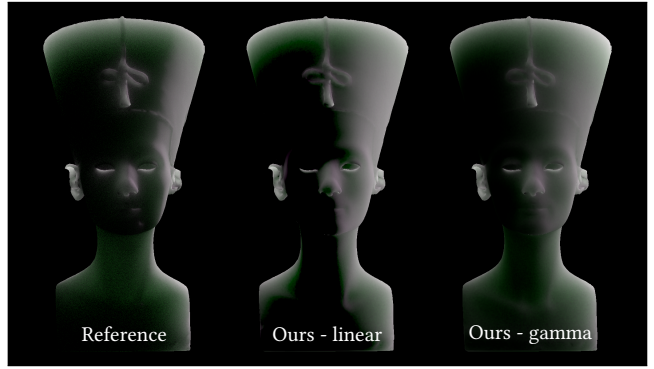

We found that for appearances with high contrast -such as subsurface scattering where details of diffusion are also to be found in low values- the reconstruction of indirect illumination was introducing high frequency contrast. To correct those artifacts, we perform the direct to indirect transfer in Gamma space. During precomputation, we Gamma-correct the values in the dataset and evaluate the basis and transfer matrix. At runtime, we Gamma-correct the incident irradiance values before applying the transfer matrix. After reconstructing the indirect illumination, we invert the Gamma correction. We show in Figure~\ref{results:nefertiti} how working in Gamma space leads to visually more acceptable results.

\subsection{Working with Animated Assets}
When working with animated assets such as rigged meshes, we construct a database of random animation keyframes with random lighting. With this database, we choose to generate a single basis that works across keyframes. Then, we can either construct a transfer matrix per keyframe or a single transfer matrix for the whole animation. We found that for subsurface scattering, this later approach worked well in practice (see our supplemental video).

%
%
\section{Results}
\label{sec_results}

We prototyped our method in the Unity game engine using the High Definition Render Pipeline~\cite{lagarde2021unity}. There, we used the Deferred Shading for all our results. We validated our method for different light transport scenarios: indirect illumination on surfaces (Section~\ref{sec:indirect_surfaces}), translucent materials (Section~\ref{sec:translucent_materials}), and multiple scattering in hair (Section~\ref{sec:indirect_hair}). We also benchmarked the performance and quality of our method with a varying number of basis components (Section~\ref{sec:performances}).

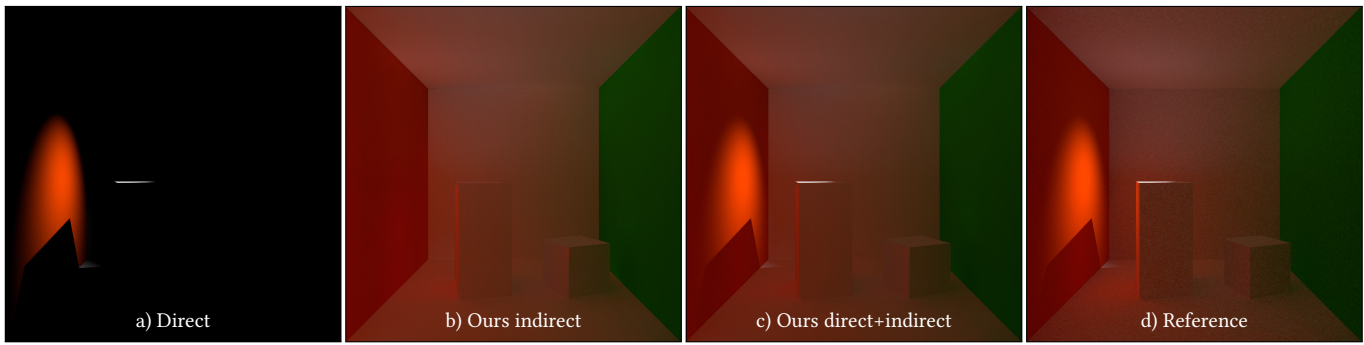
\begin{figure*}[t!]
    \begin{tikzpicture}[font=\small]
        \node[inner sep=0pt]                (A) { \includegraphics[width=0.25\linewidth,frame]{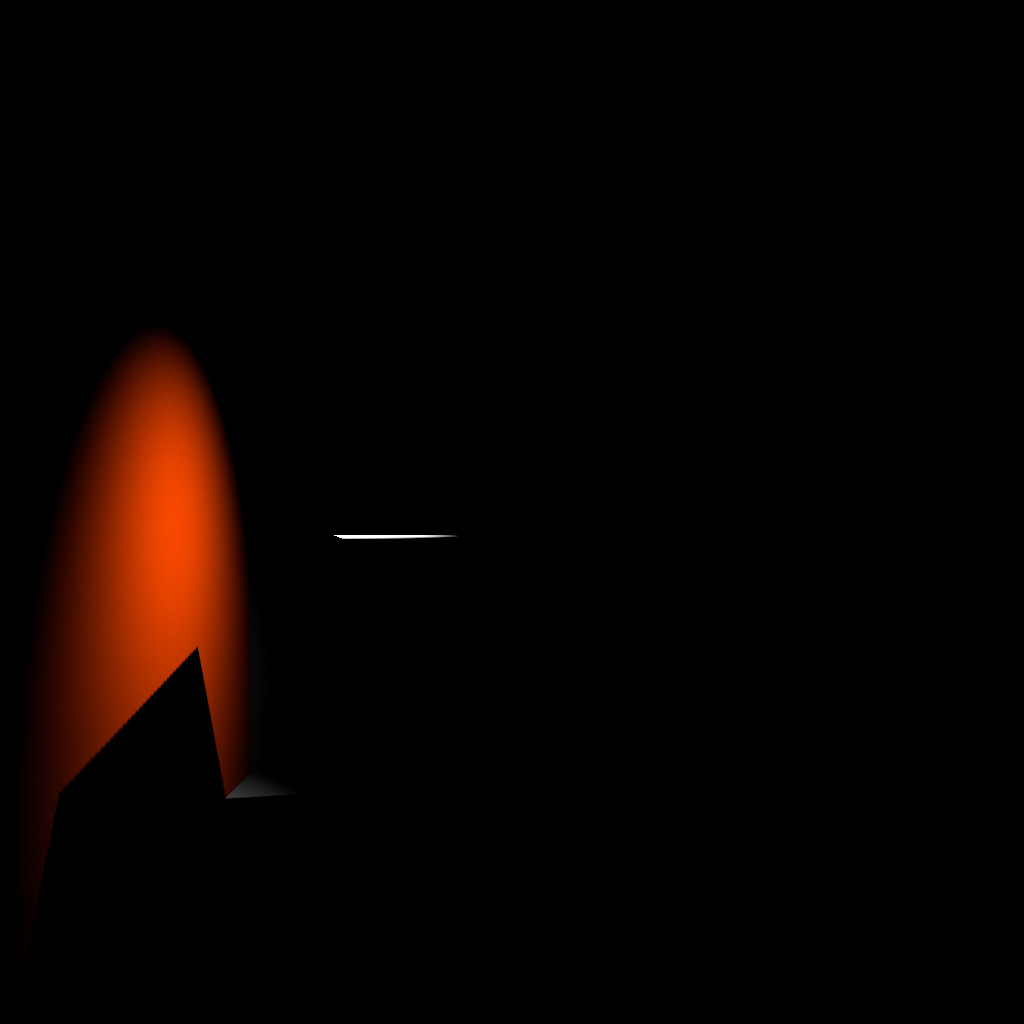} };
        \node[inner sep=0pt,right=1pt of A] (B) { \includegraphics[width=0.25\linewidth,frame]{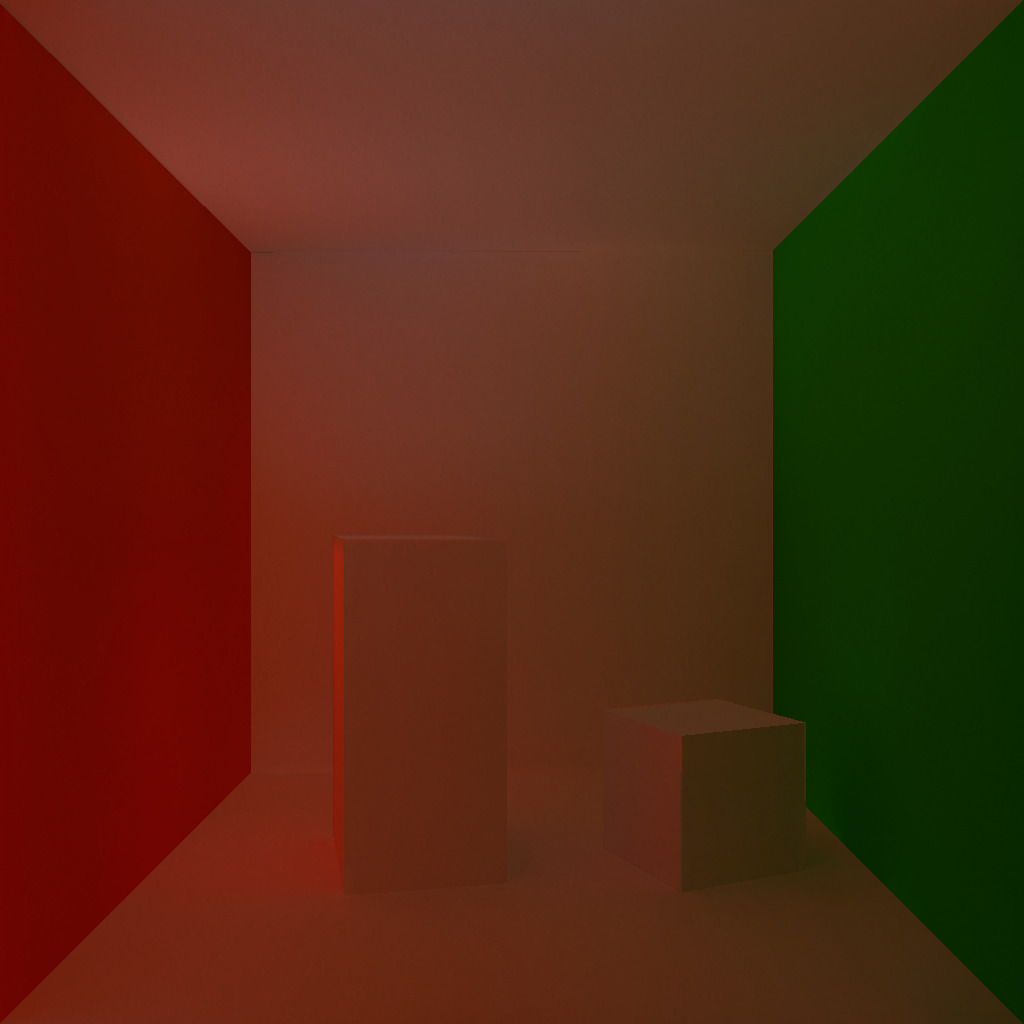} };
        \node[inner sep=0pt,right=1pt of B] (C) { \includegraphics[width=0.25\linewidth,frame]{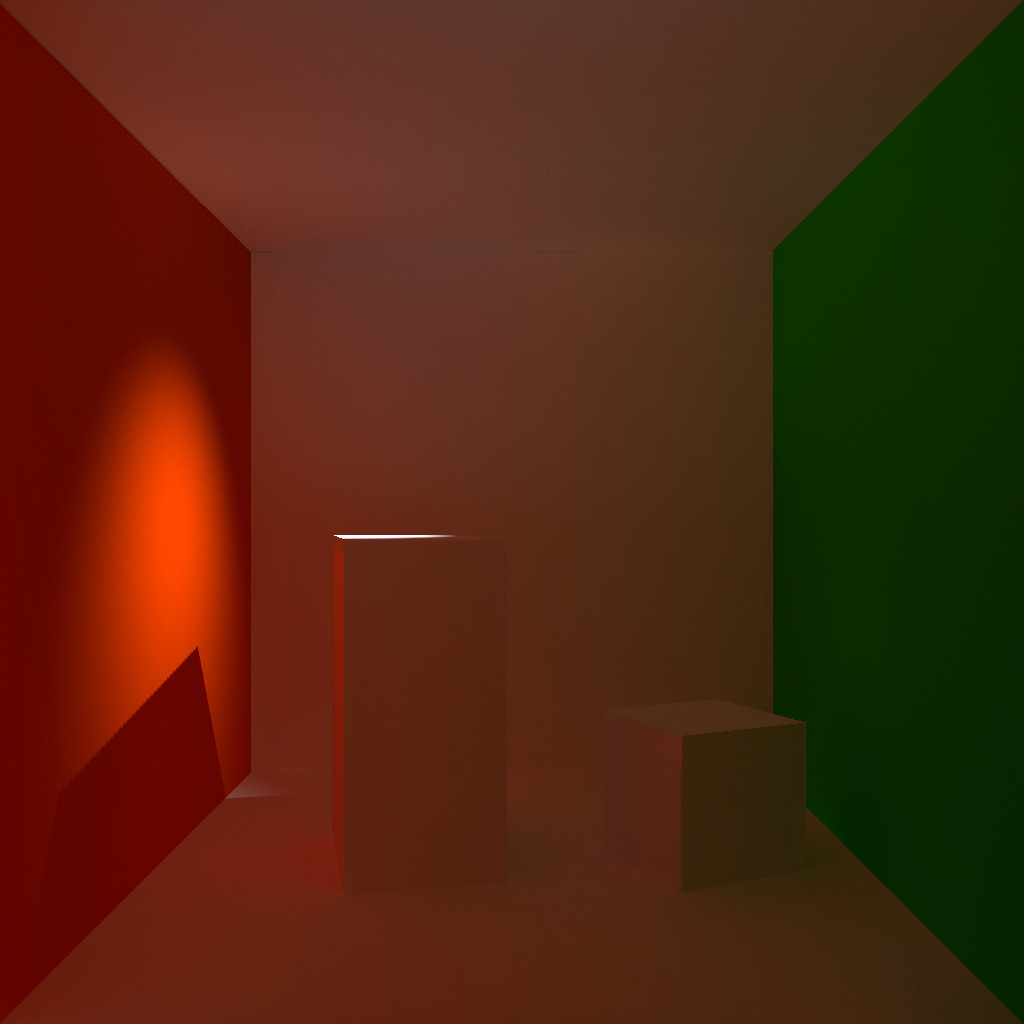} };
        \node[inner sep=0pt,right=1pt of C] (D) { \includegraphics[width=0.25\linewidth,frame]{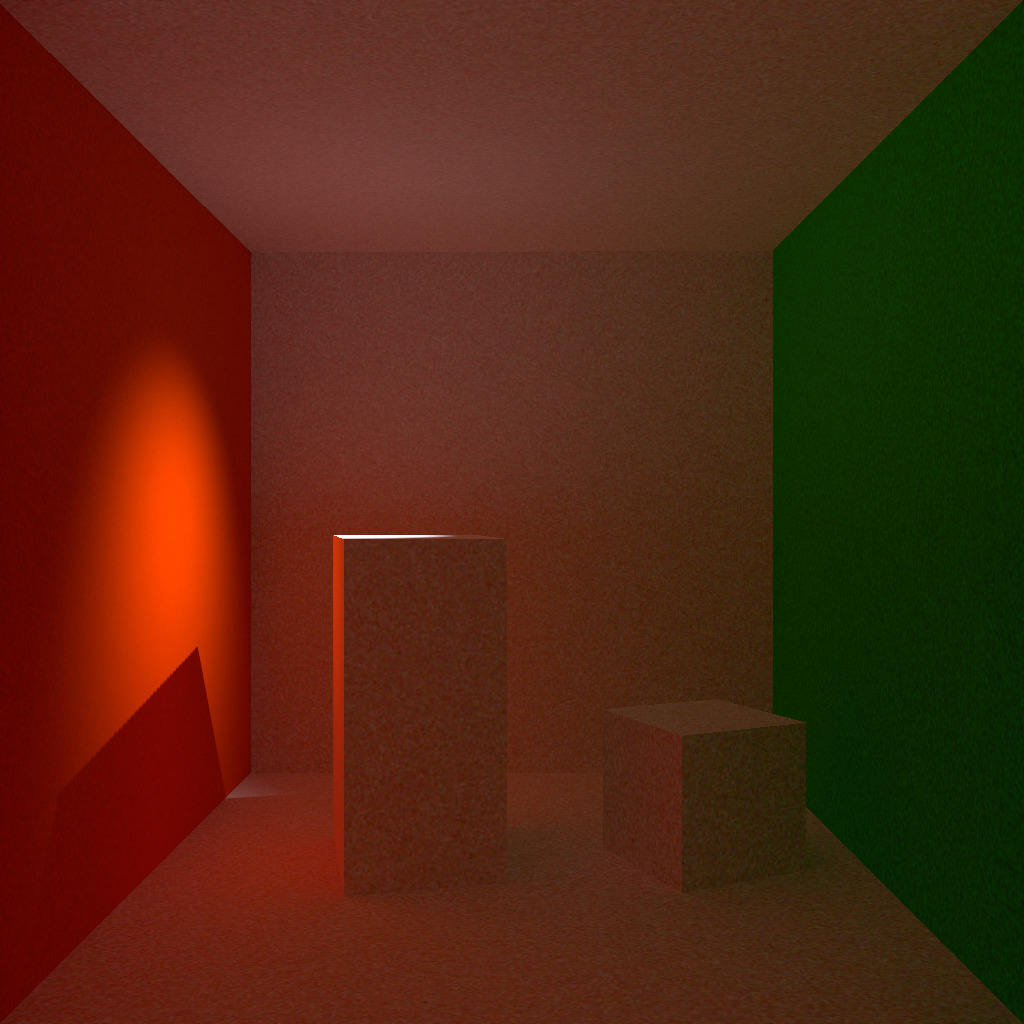} };

        \node[below=-15pt of A] {\textcolor{white}{a) Direct}};
        \node[below=-15pt of B] {\textcolor{white}{b) Ours indirect}};
        \node[below=-15pt of C] {\textcolor{white}{c) Ours direct+indirect}};
        \node[below=-15pt of D] {\textcolor{white}{d) Reference}};
    \end{tikzpicture}
    \vspace{-10pt}
    \caption{
        \textbf{Reconstructing a frame of the \textsc{Cornell Box} scene.}
        \textnormal{
            We render the direct illumination (a) using traditional techniques (in our case a Deferred Renderer). Using a custom Deferred pass on the fake G-buffer, we estimate the vector of direct illumination and reconstruct the indirect lighting by weighting the bases together with the estimated weights (b). Summing both direct and indirect together provides the final output (c). We compare it to a reference for the same lighting condition (d). In this setup, we used 128 RGB values to estimate direct illumination, 16 basis elements and a regularization of $\lambda = 0.01$.
        }
        \label{results:cornell}
    }
\end{figure*}

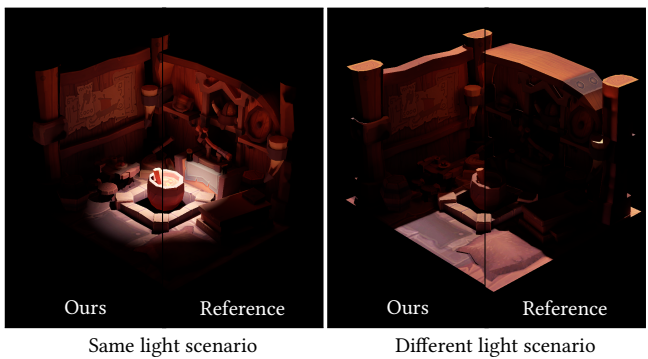
\begin{figure}[t]
    \begin{tikzpicture}[font=\small]
        \node[inner sep=0pt]               (A) { \includegraphics[width=0.25\linewidth,frame,clip,trim=0 0 512 0]{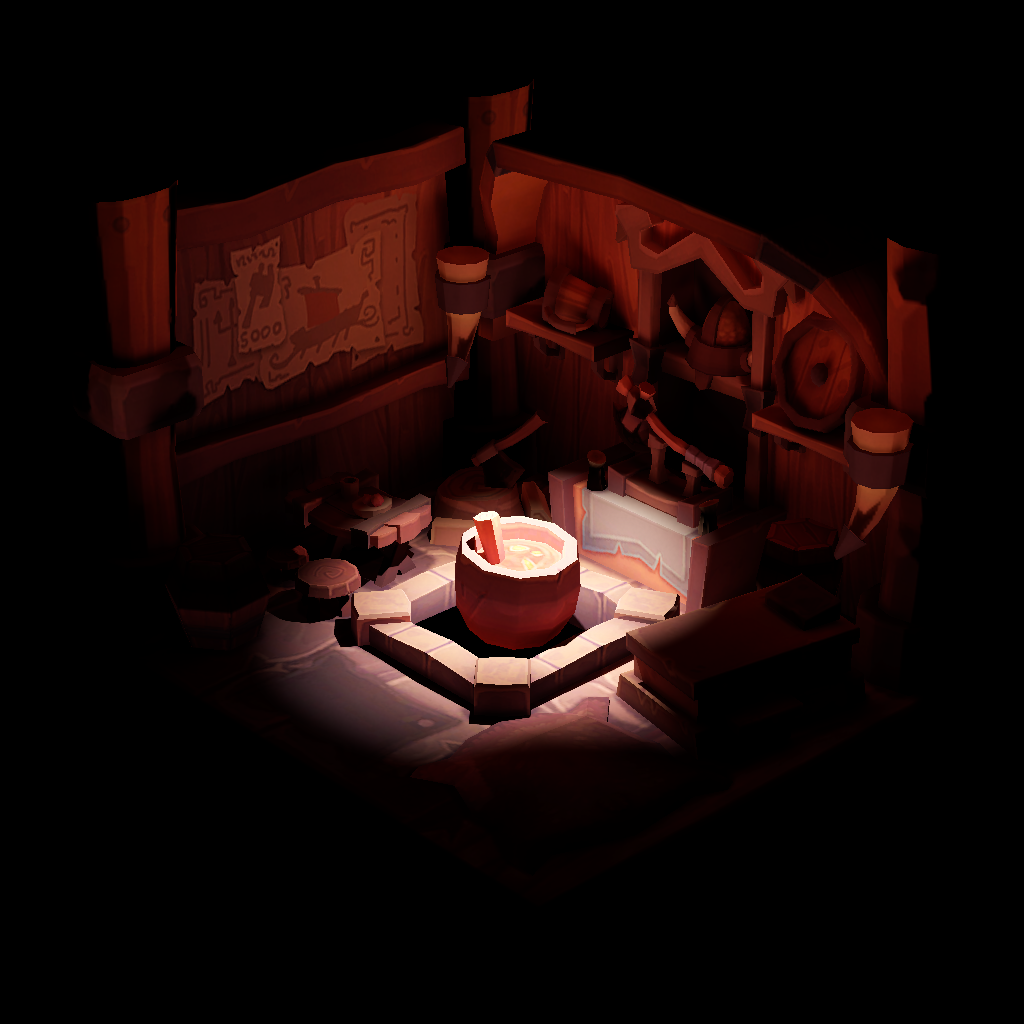} };
        \node[right=-2pt of A,inner sep=0pt] (B) { \includegraphics[width=0.25\linewidth,frame,clip,trim=512 0 0 0]{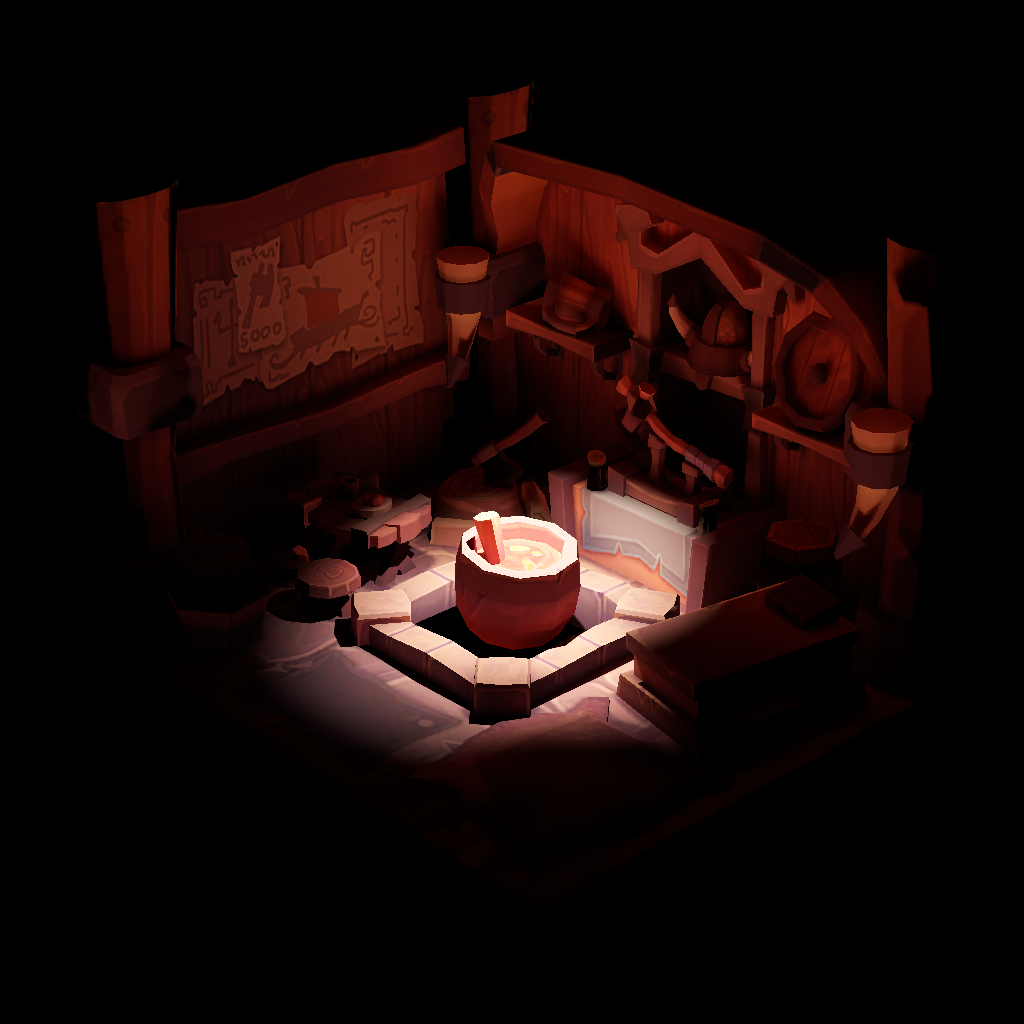} };

        \node[right=1pt of B, inner sep=0pt] (C) { \includegraphics[width=0.25\linewidth,frame,clip,trim=0 0 512 0]{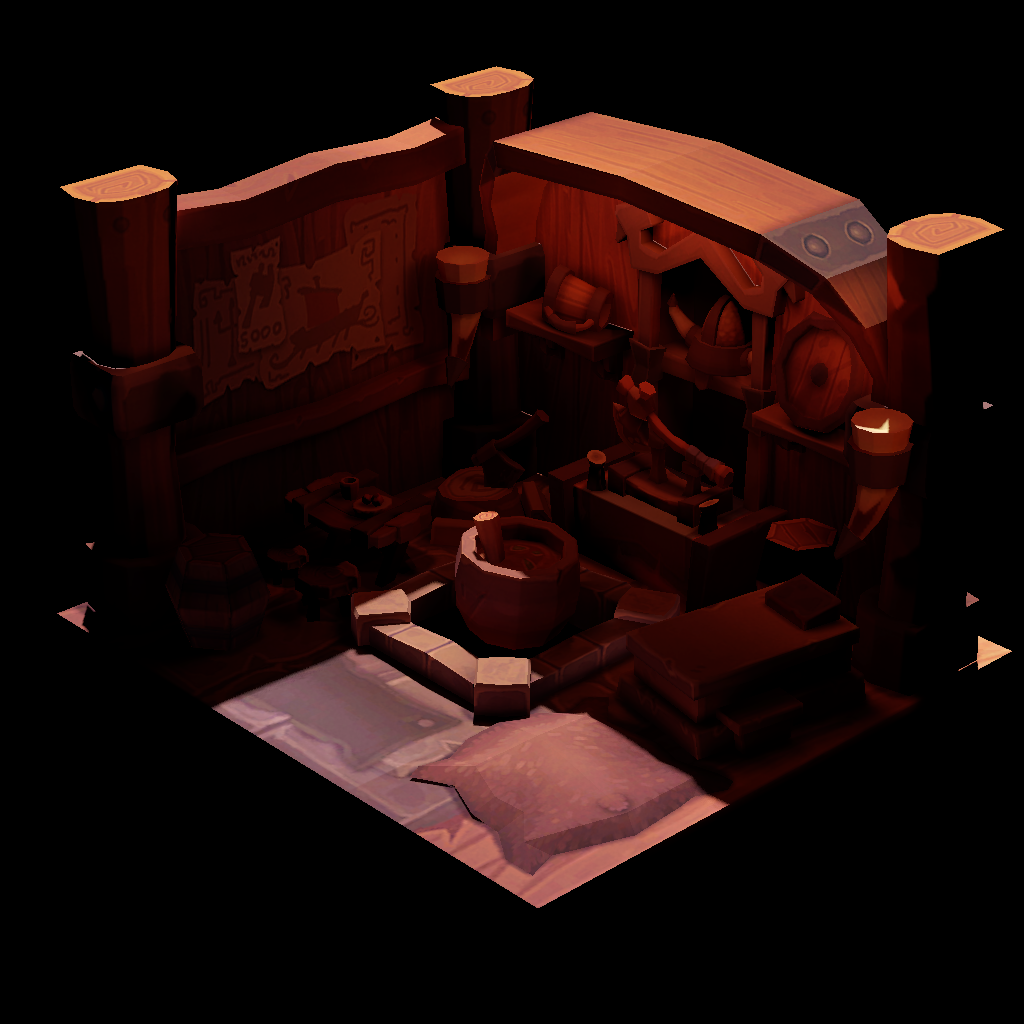} };
        \node[right=-2pt of C, inner sep=0pt] (D) { \includegraphics[width=0.25\linewidth,frame,clip,trim=512 0 0 0]{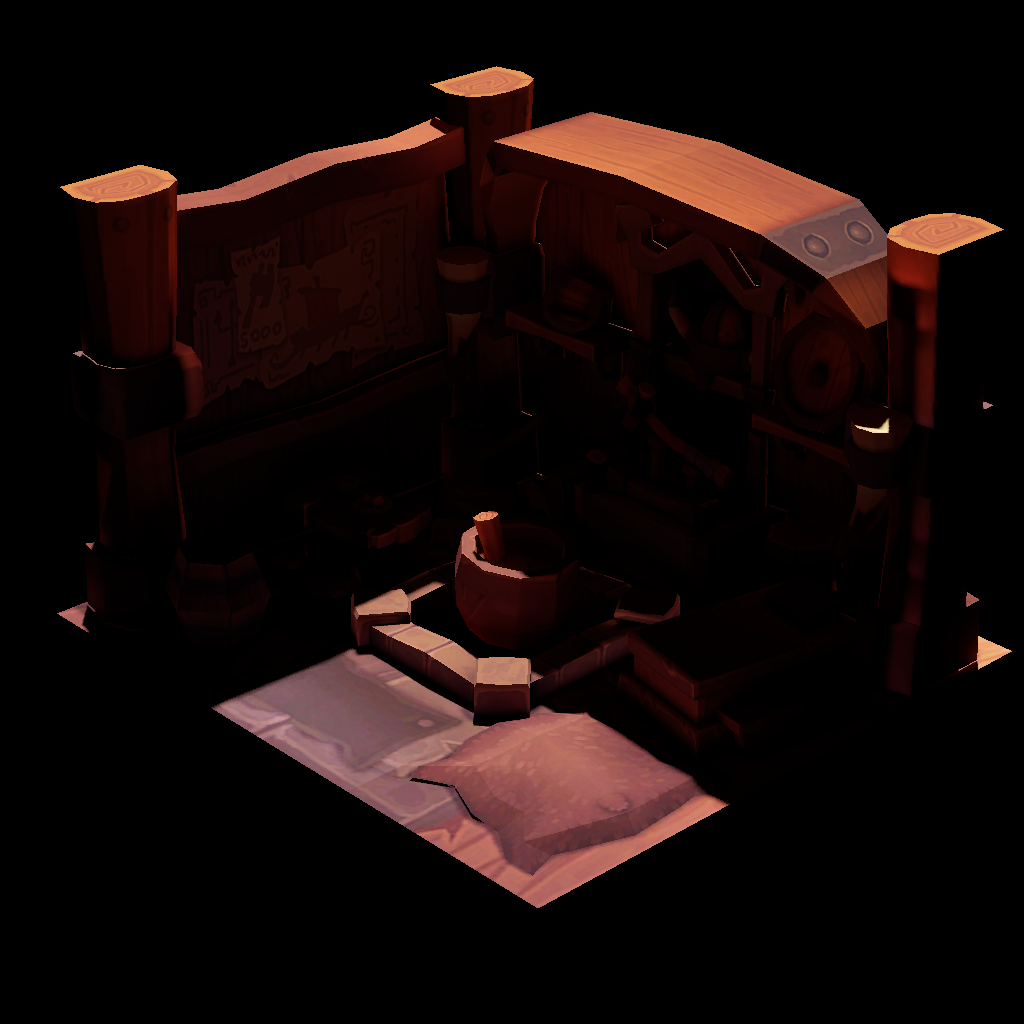} };

        \node[below=-14pt of A] {\textcolor{white}{Ours}};
        \node[below=-14pt of B] {\textcolor{white}{Reference}};
        \node[below=-14pt of C] {\textcolor{white}{Ours}};
        \node[below=-14pt of D] {\textcolor{white}{Reference}};

        \node[below=0pt of B.south east,anchor=north east,minimum width=4cm] { Same light scenario };
        \node[below=0pt of D.south east,anchor=north east,minimum width=4cm] { Different light scenario };
    \end{tikzpicture}
    \vspace{-20pt}
    \caption{
        \textbf{Changing light scenario.}
        \textnormal{
            Our reconstruction is tailored to the lighting conditions in the dataset. When we relight the scene with the same type of lighting -a spotlight- the result closely approximate the reference (left). However, when a different light condition is applied -a directional light- (right), the result, while being coherent, departs from the reference.
        }
        \label{results:viking_light}
    }
\end{figure}

\begin{figure}[t]
    \begin{tikzpicture}[font=\small]
        \node[inner sep=0pt] (A){ \includegraphics[width=0.5\linewidth,frame]{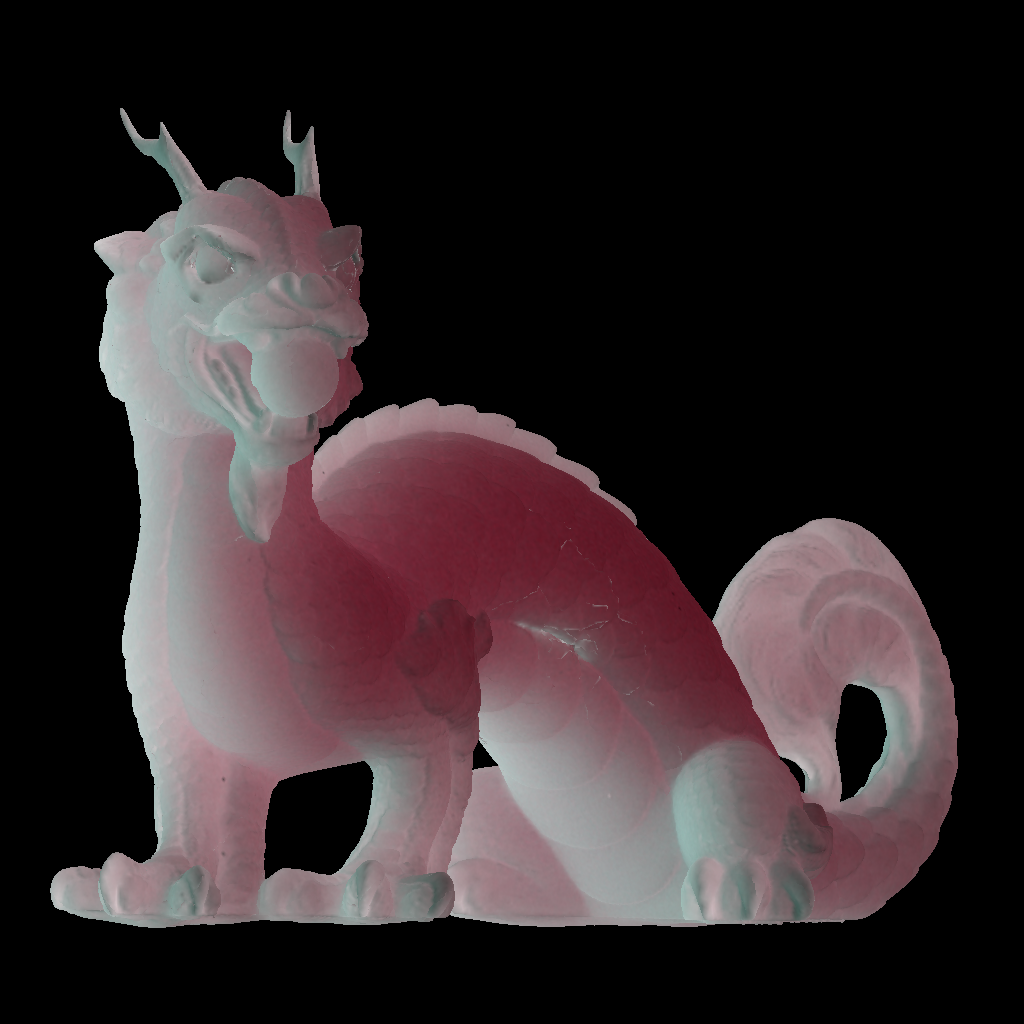} };
        \node[right=1pt of A,inner sep=0pt] (B){ \includegraphics[width=0.5\linewidth,frame]{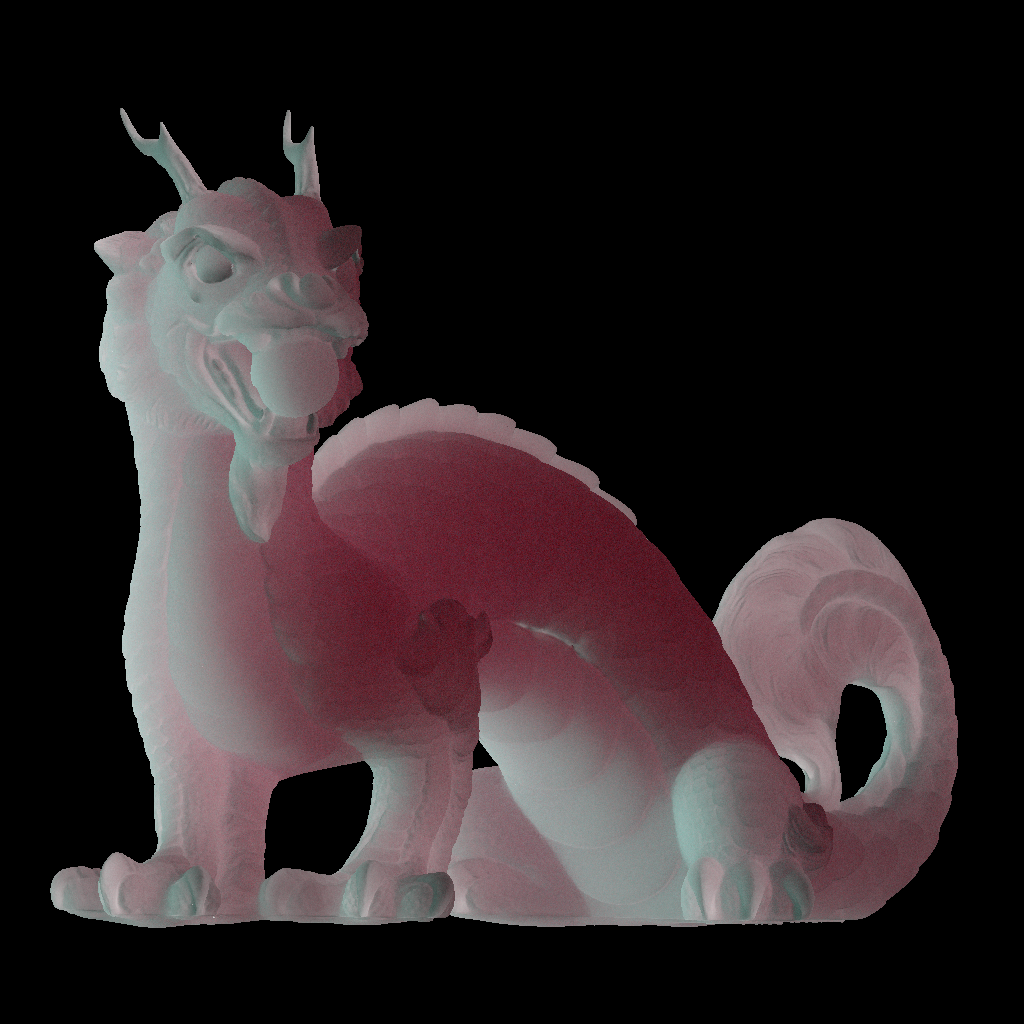} };

        \node[below=-13pt of A] {\textcolor{white}{Ours}};
        \node[below=-13pt of B] {\textcolor{white}{Reference}};
    \end{tikzpicture}
    \vspace{-10pt}
    \caption{
        \textbf{Rendering Subsurface scattering.}
        \textnormal{
            Our method enables to render translucency due to thick participating media in real-time. It handles light transmission well, something that screen-space methods struggle to cope with.
        }
        \label{results:dragon}
        \vspace{-10pt}
    }
\end{figure}

\begin{figure}[t]
    \begin{tikzpicture}[font=\small]
        \node[inner sep=0pt]               (C) { \includegraphics[width=0.5\linewidth,frame]{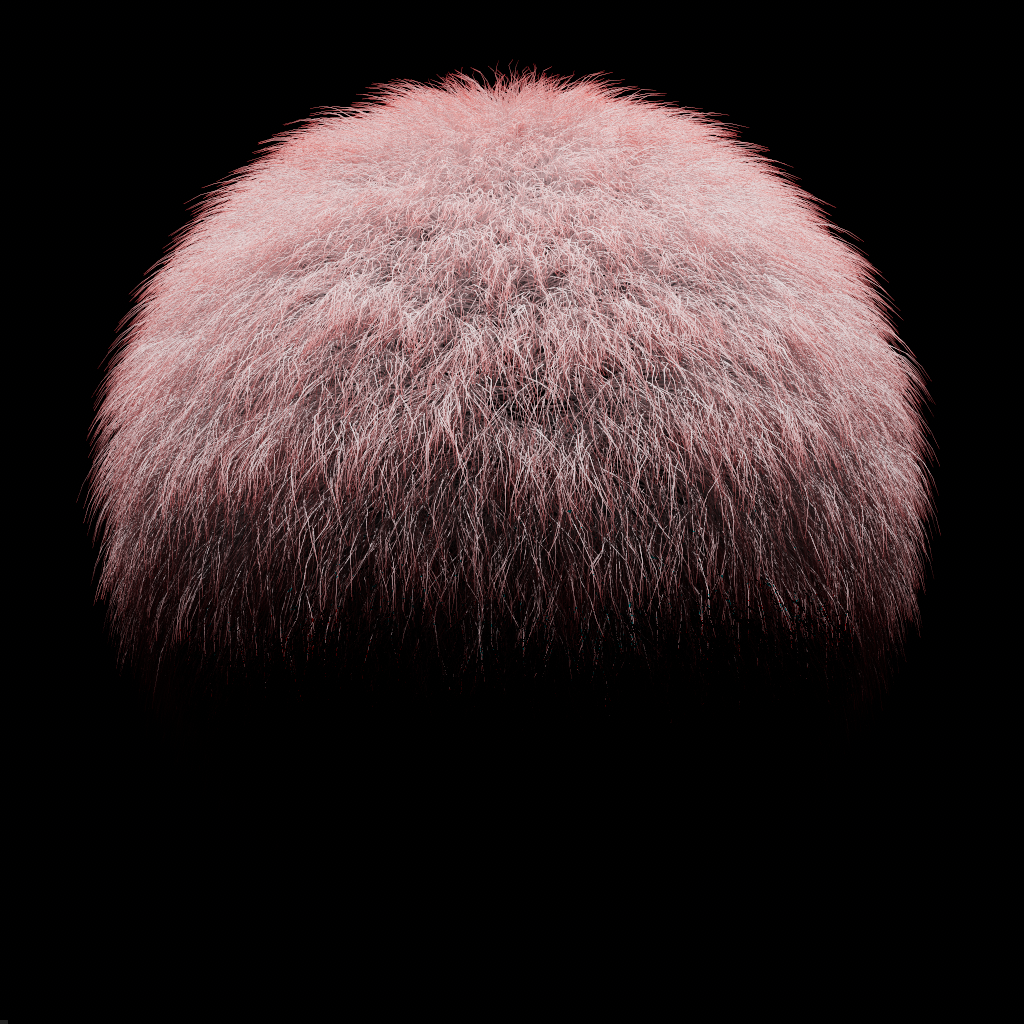} };
        \node[inner sep=0pt,right=1pt of C] (D) { \includegraphics[width=0.5\linewidth,frame]{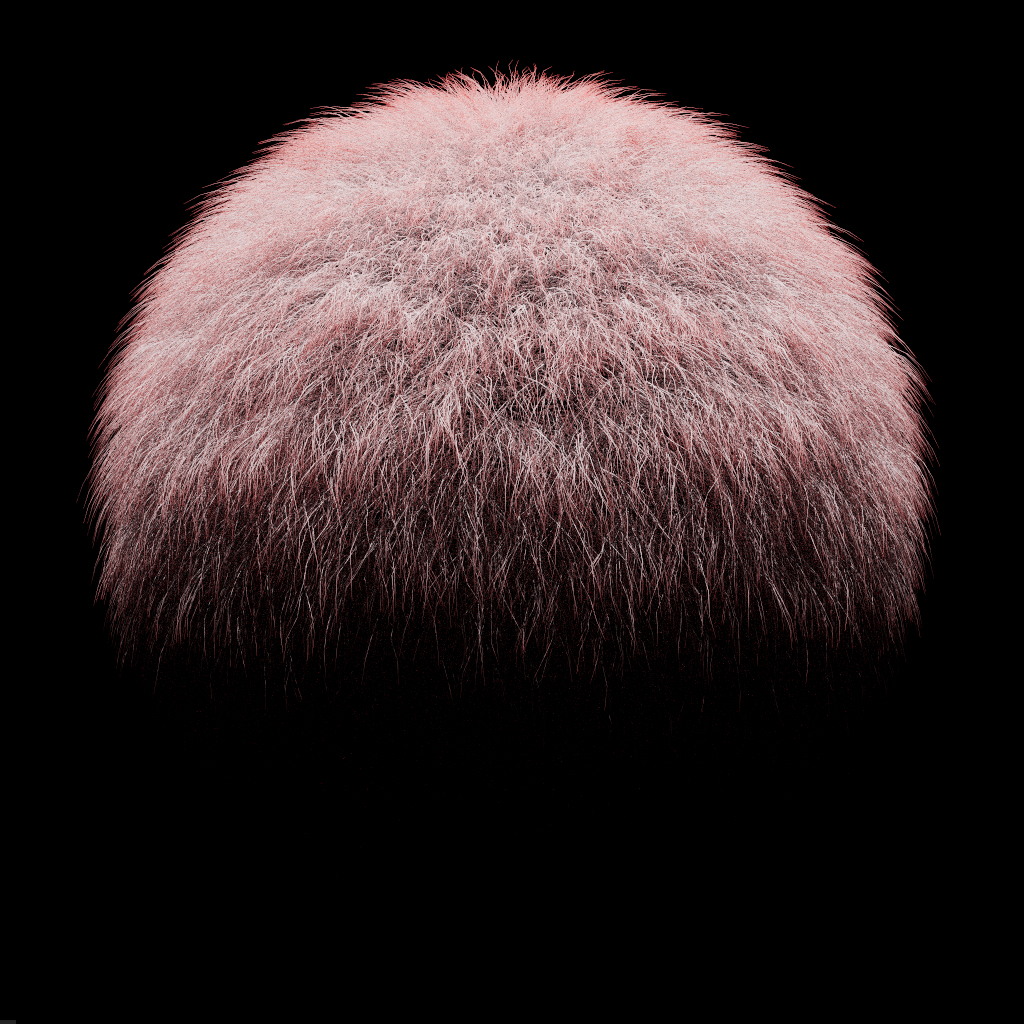} };

        \node[below=-13pt of C] {\textcolor{white}{Ours}};
        \node[below=-13pt of D] {\textcolor{white}{Reference}};
    \end{tikzpicture}
    \vspace{-15pt}
    \caption{
        \textbf{Hair rendering.}
        \textnormal{
            We render the interreflection of light in this \textsc{Hairball} asset containing $140\mbox{k}$ hair strands. The direct illumination is evaluated on the hull of the hair volume. We store the basis for the indirect lighting in textures using $10$ texels per fiber (see Figure~\ref{fig:hair_parameterization}). 
        }
        \label{results:hairball}
    }
\end{figure}

\begin{figure}[t]
    \begin{tikzpicture}[font=\small]
        
        \node[inner sep=0] (A)  { \includegraphics[height=4cm,frame]{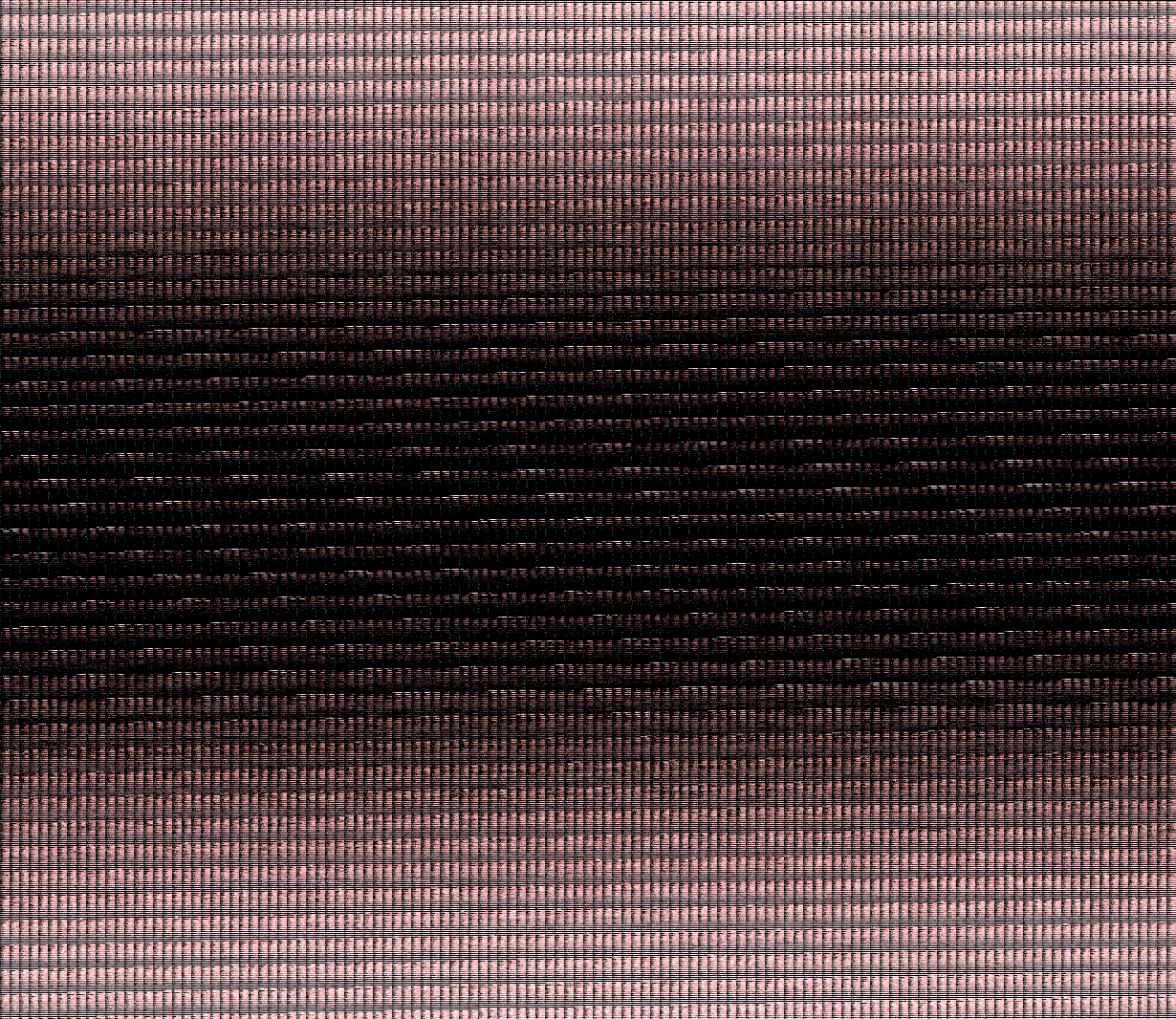} };
        \draw[color=red] (A.south west) + (0.8cm,0.96cm) rectangle +(1.5cm,1.6cm);
        \node[inner sep=0, right=2pt of A] (B)  { \includegraphics[height=4cm,frame,clip,trim=256 256 718 544]{./figures/hairball/parameterization.png} };
        \draw[color=red,line width=1pt] (B.south west) rectangle (B.north east);
    \end{tikzpicture}
    \vspace{-15pt}
    \caption{
        \textbf{Hair parameterization.}
        \textnormal{
            We parameterize indirect illumination in hair using hair index and arc length that we pack using 10 texels per strands (see inset on the right). In this figure, we display one element of the dataset of the \textsc{Hairball} asset. For each texel, we average the outgoing radiance along the hair's azimuth as well as along the arc length inside the texel.
        }
        \label{fig:hair_parameterization}
    }
\end{figure}

\subsection{Indirect Illumination on Surfaces}
\label{sec:indirect_surfaces}
With the \textsc{Cornell Box} scene in Figure~\ref{results:cornell} we showcase the use of our method to render Global Illumination in a static scene. Because our basis provides a global support for the indirect illumination, we found that it is not a good fit for large scene GI as artifacts are likely to appear. Precomputed Radiance Transfer here would tessellate the scene to perform local reconstruction of indirect lighting resulting in fewer artifacts but requiring more computational power. \\

With the \textsc{Viking} scene in Figure~\ref{results:viking_light}, we show how our method is tailored to a specific lighting scenario. In this figure, we change the spot lighting used to build the dataset to point lights close to the surface of the mesh. While the resulting indirect illumination looks coherent when moving the point lights, the reconstructed result departs from the ground truth. This validates that our basis decomposition is optimized for specific lighting conditions.

\subsection{Subsurface Scattering}
\label{sec:translucent_materials}
Our method supports diffusion in thick media enclosed in a geometry. \textsc{Lizard}, \textsc{Nefertiti}, and \textsc{Dragon} scenes (see Figure~\ref{fig:teaser}, \ref{results:nefertiti} and~\ref{results:dragon} respectively) demonstrate its use in such context. Our method captures the self shadowing and the color saturation due to multiple scattering within the object. As shown in Figure~\ref{results:nefertiti}, the rendering of subsurface scattering creates high contrasts that do not reconstruct well in linear space. By applying a Gamma correction during the direct to indirect transfer, we get visually closer results.

\subsection{Scattering in Hair}
\label{sec:indirect_hair}
We also use our method to render indirect illumination in hair as demonstrated in scenes \textsc{Lizard} (Figure~\ref{fig:teaser}) and \textsc{Hairball} (Figure~\ref{results:hairball}). Unlike rendering on a mesh we evaluate direct illumination on the hull defined by hair fibers. The extraction of the indirect illumination basis follows the same process, however the texture-space parameterization of hair is different.

\paragraph{Texture Space for Hair} When using hair, we parameterize hair strands using length and strand index as 2D coordinates. We further pack this texture coordinate by using 10 pixels per hair strand and concatenating multiple strands per row. Using this packing, we have $100\mbox{k}$ hair strands in a $1000 \times 1000$ texture (see Figure~\ref{fig:hair_parameterization}). The \textsc{Lizard} and the \textsc{Hairball} have respectively $10k$ and $140k$ hair strands. In our results, we reconstruct a radially constant illumination at each point on the fiber. While directional variation would increase the accuracy at the expense of additional storage, we found that it was not visually necessary on our assets. 

\begin{figure}[h]
    \begin{tikzpicture}[font=\small]
        \node[inner sep=0pt] (A) { \includegraphics[width=0.33\linewidth]{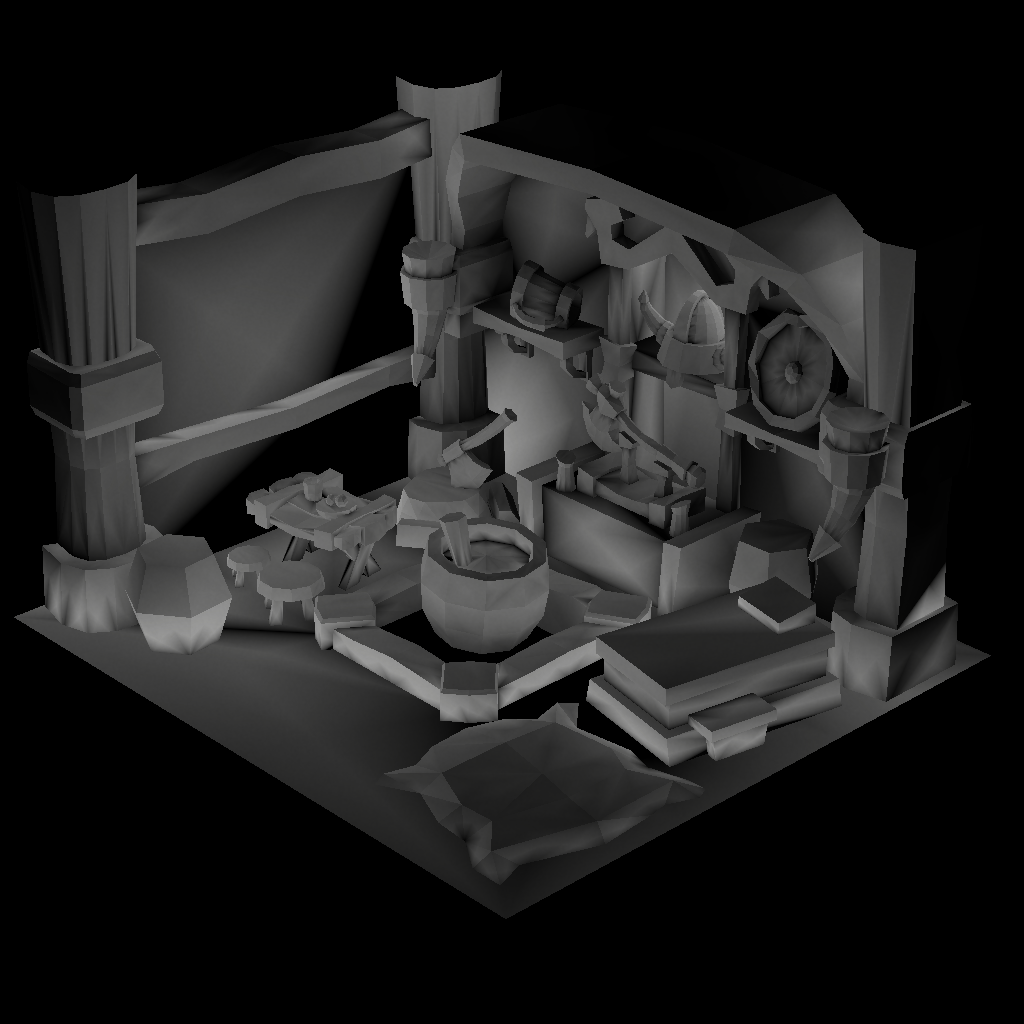} };
        \node[inner sep=0pt, right=-1pt of A] (B) { \includegraphics[width=0.33\linewidth]{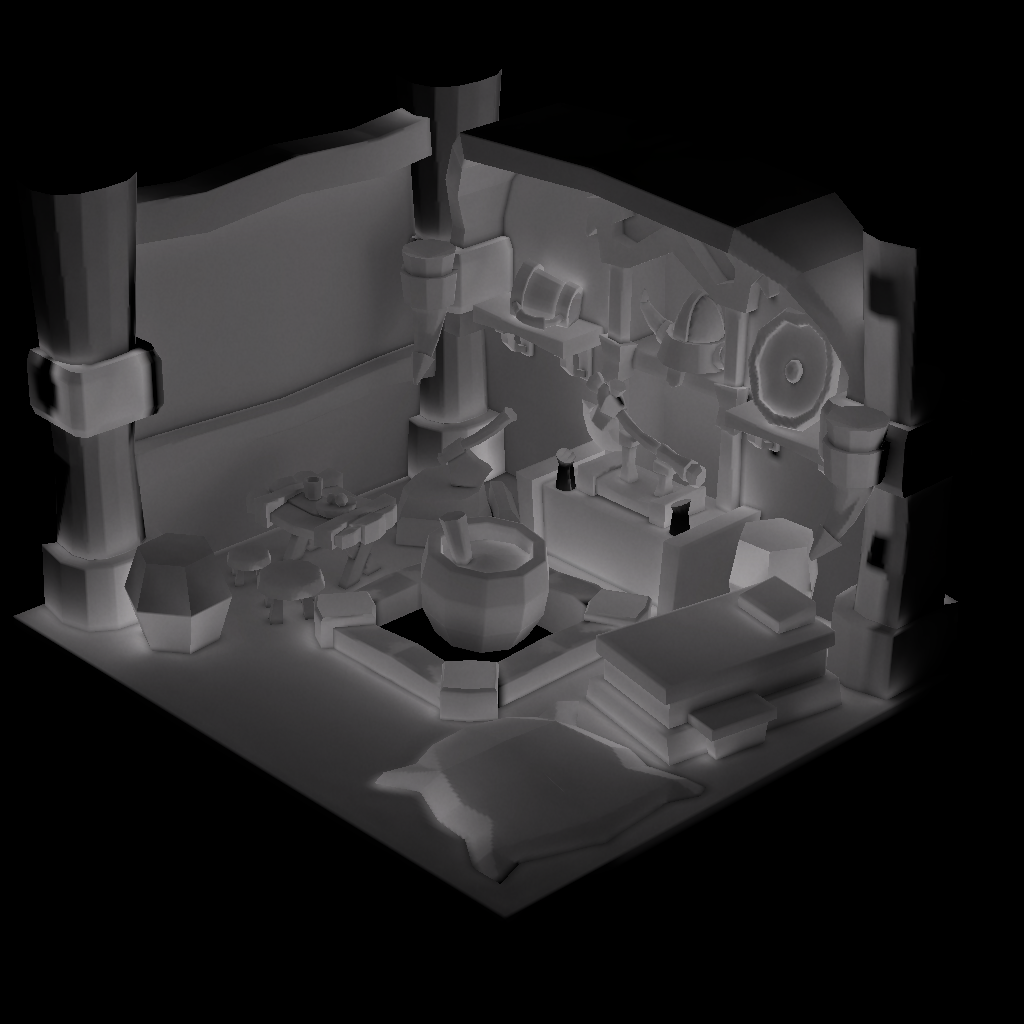} };
        \node[inner sep=0pt, right=-1pt of B] (C) { \includegraphics[width=0.33\linewidth]{./figures/prt/cmp_prt_ref.png} };

        \node[below=0pt of A, anchor=south] { \textcolor{white}{Classical PRT} };
        \node[below=0pt of B, anchor=south] { \textcolor{white}{Reference} };
        \node[below=0pt of C, anchor=south] { \textcolor{white}{Ours} };
    \end{tikzpicture}
    \vspace{-7pt}
    \caption{
        \update{
        \textbf{Comparing to Classical PRT.}
        \textnormal{
            We compare our method with a classical PRT method~\protect\cite{sloan2002precomputed}. While classical PRT both blur the incident illumination and is linked to the tessellation of the asset, our method is free of those issues.
        }
        }
        \label{fig:cmp_prt}
        \vspace{-7pt}
    }
\end{figure}

\update{
\subsection{Comparison with Classical PRT}
\label{sec:cmp_prt}
We compare our method with classical PRT~\cite{sloan2002precomputed} to render indirect illumination from surfaces in the \textsc{Lion} scene lit by a directional light (in Figure~\ref{fig:cmp_prt}). There, we highlight some key differences with our work. First, because classical PRT restricts the frequency content of the incoming lighting, we can see that the directional light \textit{leaks} behind the object. Our method does not restrict the frequency content of incoming light but rather the space of possible indirect illumination. Hence, we can better reproduce such lighting scenario. Furthermore, classical PRT is performed on the vertices of the asset. This can cause interpolation artifacts when the asset is poorly tessellated, and it also links performance to the vertex count. Since we rely on a meshless approach, we are free of issues.
}

\subsection{Performance, Storage, \& Quality}
\label{sec:performances}
We captured the performance of our method on an Nvidia RTX 2080 and compared it to Unity's standard shader with no indirect illumination as a baseline. We report the timings when using 4, 16, 64 basis elements (respectively 1, 4, and 16 RGBA textures) as well as baking times for a dataset of $128$ elements in Table~\ref{tbl:performances}. \update{The rendering time is decomposed into: shader evaluation; fake G-buffer computation; and matrix transfer to compute the indirect illumination coefficients. We found that this later part is independent of the number of basis elements in the 3 tested configurations.}

\begin{table}[h]
    \caption{
        \label{tbl:performances}
        \update{
        \textbf{Performance.}    
        \textnormal{
            We compare our method to a Standard shader with no indirect illumination. For our baseline, we report timings of the final fragment shader, as well as, fake G-buffer rendering (GPU), coefficients evaluation (GPU), and baking time (CPU).
        }
        }
        \vspace{-17pt}
        }
    \begin{tabular}{|c |c c c c|} 
        \hline
                           & \textsc{Cornell} & \textsc{Viking} & \textsc{Dragon} & \textsc{Hairball} \\
        \hline\hline
        No GI              & $221 \,\mu s$    & $232\,\mu s$    & $418\,\mu s$    & $8729\,\mu s$  \\
        \hline \hline
        Ours (4 basis)     & $325 \,\mu s$    & $348\,\mu s$    & $490\,\mu s$    & $8910\,\mu s$  \\ 
        Ours (16 basis)    & $428 \, \mu s$   & $483\,\mu s$    & $658\,\mu s$    & $9547\,\mu s$  \\ 
        Ours (64 basis)    & $892 \,\mu s$    & $1068\,\mu s$   & $1358\,\mu s$   & $11971\,\mu s$ \\ 
        \hline \hline
        Fake G-buffer      & $36\,\mu s$      & $34\,\mu s$     & $32\,\mu s$    & $34\, \mu s$   \\
        Matrix transfer    & $15\,\mu s$      & $15\,\mu s$     & $15\,\mu s$    & $15\,\mu s$   \\
        \hline \hline
        Bake time          & $6.48\,min$      & $6.89\,min$     & $13.37\,min$    & $49.07\,min$   \\
        \hline
    \end{tabular}
\end{table}

The storage footprint for a $16$ monochannel basis at $1024 \times 1024$ resolution with $32$ bits per channel is $64$MB on GPU (4 RGBA floating point textures) and the associated transfer matrix for $256$ measurement points weighs $36$KB. \\

\begin{figure}[b]
    \begin{tikzpicture}[font=\small]
        \node[inner sep=0pt] (A) { \includegraphics[width=\linewidth,frame]{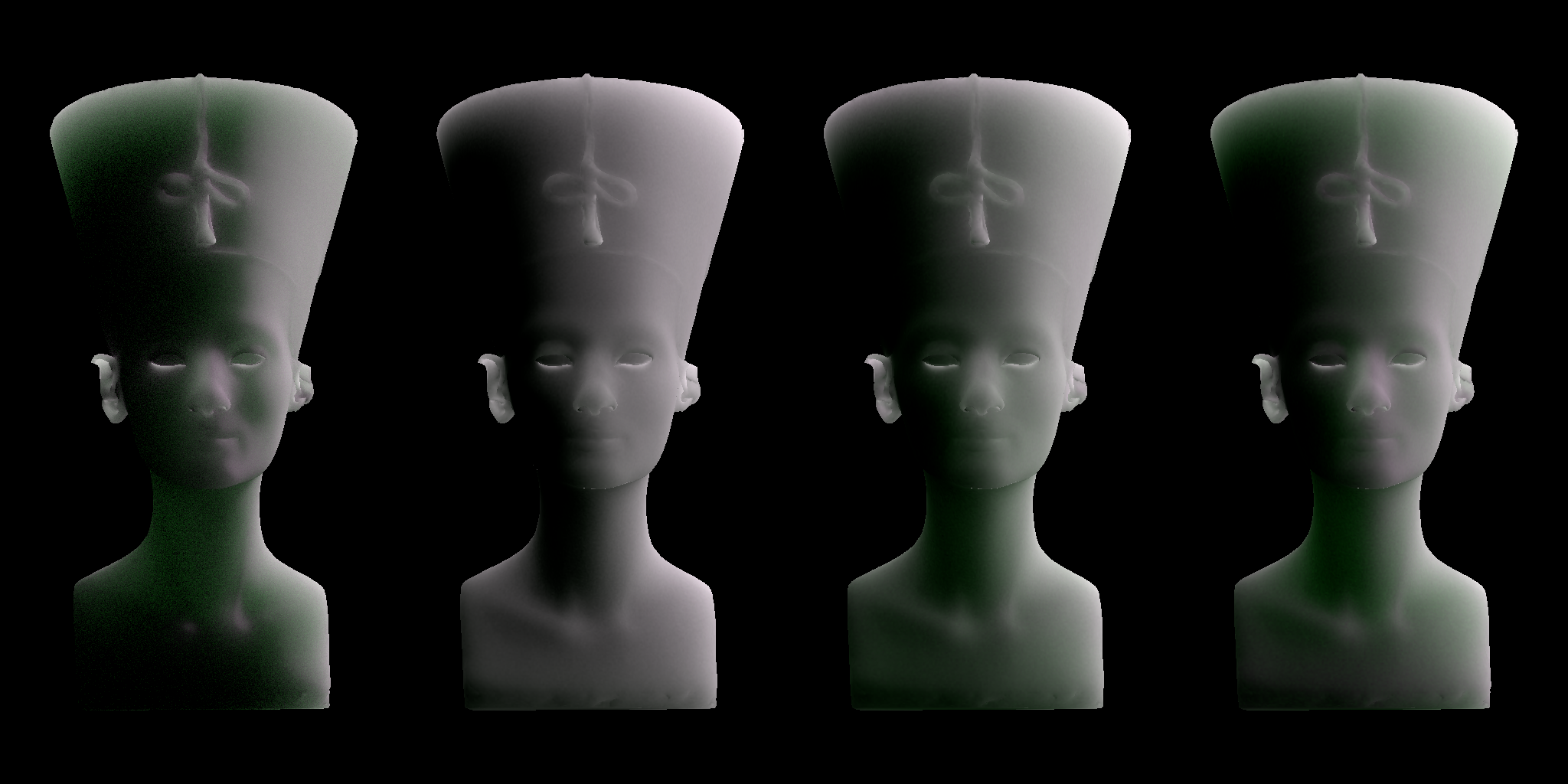} };
        \node[right=0.5cm of A.south west, anchor=south west] { \textcolor{white}{Reference} };
        \node[right=2.8cm of A.south west, anchor=south west] { \textcolor{white}{$4$ bases} };
        \node[right=4.8cm of A.south west, anchor=south west] { \textcolor{white}{$16$ bases} };
        \node[right=6.9cm of A.south west, anchor=south west] { \textcolor{white}{$64$ bases} };
    \end{tikzpicture}
    \vspace{-17pt}
    \caption{
        \textbf{Increasing the number of basis.}
        \textnormal{
            The quality of the reconstruction increases as we add more basis elements. We found that for most of our assets, using $64$ basis elements (hence, $16$ textures) would provide accurate results. However, using $16$ basis elements ($4$ texture) already gives plausible results.
        }
        \label{fig:progression_basis}
        \vspace{-5pt}
    }
\end{figure}
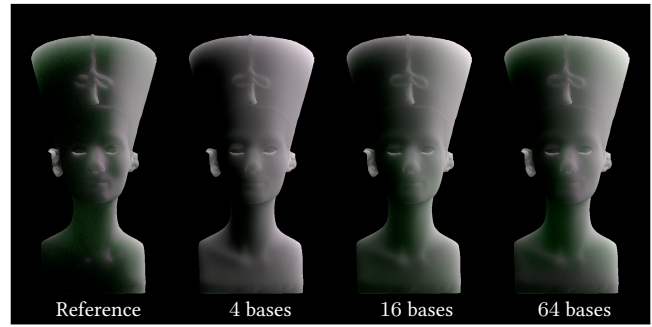

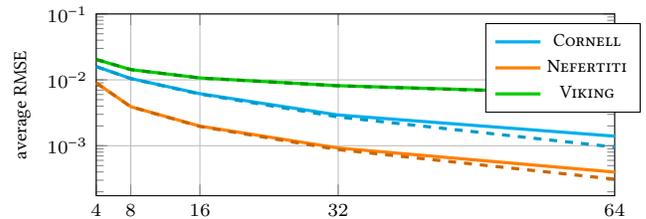
\begin{figure}[b]
    \begin{tikzpicture}[font=\footnotesize]
        \begin{semilogyaxis}[
            grid=major,
            width = \linewidth,
            height = 4.0cm,
            xmin = 4.0, xmax = 64.0,
            ymin = 0.0, ymax = 0.1,
            ylabel = {average RMSE},
            xtick = {4,8,16,32,64,128},
            legend pos = north east,
            legend style= {
                    at={(1.05, 0.95)},
                    nodes= {
                        scale=1.0,
                        transform shape
                    }
                },
            ]
            \addplot[black!00!cyan,line width=1.2pt] table {figures/quality/cornell_ours.txt};
            \addlegendentry{\textsc{Cornell}}
            \addplot[black!20!cyan,dashed,line width=1.2pt,forget plot] table {figures/quality/cornell_pca.txt};
            %
            \addplot[black!00!orange,line width=1.2pt] table {figures/quality/nefertiti_ours.txt};
            \addlegendentry{\textsc{Nefertiti}}
            \addplot[black!20!orange,dashed,line width=1.2pt,forget plot] table {figures/quality/nefertiti_pca.txt};
            %
            \addplot[black!20!green,line width=1.2pt] table {figures/quality/viking_ours.txt};
            \addlegendentry{\textsc{Viking}}
            \addplot[black!40!green,dashed,line width=1.2pt,forget plot] table {figures/quality/viking_pca.txt};
        \end{semilogyaxis}
    \end{tikzpicture}
    \vspace{-10pt}
    \caption{
        \label{fig:quality}
        \textbf{Quality.}
        \textnormal{
            The RMSE of our method w.r.t. the reference  (plain) decreases with the number of basis (abscissa), close to the RMSE of projecting the reference in the basis (dashed).
        }
    }
\end{figure}

Quality naturally increases as we grow the number of basis elements (see Figure~\ref{fig:progression_basis}). Furthermore, we found that using only a few direct illumination samples as input produces a reconstruction error close to projecting the complete indirect illumination in the basis. We report the RMSE of both approaches in Figure~\ref{fig:quality}. For this particular test, we set the regularization to $\lambda = 10^{-3}$.

%
%
\section{Limitations \& Future Work}
\update{
\paragraph{Sparse Illumination Measurement}
As shown in Section~\ref{sec:direct_vectors}, the sampling of the measurement points is linked to the achievable lighting dimensionality. Thus, it needs to be sufficiently dense to reproduce the space of observable lighting configurations. It follows that a lighting scenario mixing many light types might require a denser sampling.

\paragraph{No Directionality} We reconstruct a diffuse appearance when reconstructing indirect illumination. However, since our method does not depend on the encoding of the measured indirect illumination,
it can be extended to reconstruct glossy appearances e.g. directional distributions using directional sampling or any basis such as Spherical Harmonics. However, our method is likely to be restricted to low frequency gloss here and will not work to render specular reflections.
}

\paragraph{Large Assets} Our solution is not designed to handle assets such as levels in a game. Because we handle light transport globally and reduce it with a handful of basis functions, we cannot reconstruct the interconnected interiors or large environments in which the combinatorics of possible illumination is large. For such case, our method would require to be extended to handle modular transfer between disjoint transport solutions (Similar to~\citet{loos2011modular}).

\section{Conclusion}
We have presented a new data-driven paradigm for Precomputed Radiance Transfer methods. We have shown that it could lead to a lean and efficient method that uses images of direct and indirect lighting obtained from renderers such as path tracers to compute a transfer matrix and a reconstruction basis that are tailored to render the indirect lighting of a lighting scenario in real-time with a small footprint. This method applies to the rendering of indirect illumination from surfaces, subsurface scattering, or hair rendering and could serve as a baseline for future work in Machine Learning.

\begin{acks}
The authors thank Eric Heitz, Emmanuel Turquin, and Sebastien Largarde for discussions during the early phase of this project.
The \textsc{Lizard} scene contains models from Marleen Vijgen and Conrad Justin. The \textsc{Viking} room was modeled by Niegel Goh. The \textsc{Dragon} was scanned by Arkify. The \textsc{Hairball} was generated by Jonathan Dupuy.
\end{acks}

\vfill

%
%

\bibliographystyle{ACM-Reference-Format}
\bibliography{bibliography}


\begin{thebibliography}{28}


\ifx \showCODEN    \undefined \def \showCODEN     #1{\unskip}     \fi
\ifx \showDOI      \undefined \def \showDOI       #1{#1}\fi
\ifx \showISBNx    \undefined \def \showISBNx     #1{\unskip}     \fi
\ifx \showISBNxiii \undefined \def \showISBNxiii  #1{\unskip}     \fi
\ifx \showISSN     \undefined \def \showISSN      #1{\unskip}     \fi
\ifx \showLCCN     \undefined \def \showLCCN      #1{\unskip}     \fi
\ifx \shownote     \undefined \def \shownote      #1{#1}          \fi
\ifx \showarticletitle \undefined \def \showarticletitle #1{#1}   \fi
\ifx \showURL      \undefined \def \showURL       {\relax}        \fi
\providecommand\bibfield[2]{#2}
\providecommand\bibinfo[2]{#2}
\providecommand\natexlab[1]{#1}
\providecommand\showeprint[2][]{arXiv:#2}

\bibitem[\protect\citeauthoryear{Abrash}{Abrash}{2000}]%
        {abrash2000quake}
\bibfield{author}{\bibinfo{person}{Michael Abrash}.}
  \bibinfo{year}{2000}\natexlab{}.
\newblock \showarticletitle{Quake’s lighting model: Surface caching}.
\newblock \bibinfo{journal}{\emph{Graphic programming black book}}
  (\bibinfo{year}{2000}).
\newblock


\bibitem[\protect\citeauthoryear{Bitterli, Wyman, Pharr, Shirley, Lefohn, and
  Jarosz}{Bitterli et~al\mbox{.}}{2020}]%
        {bitterli2020spatiotemporal}
\bibfield{author}{\bibinfo{person}{Benedikt Bitterli}, \bibinfo{person}{Chris
  Wyman}, \bibinfo{person}{Matt Pharr}, \bibinfo{person}{Peter Shirley},
  \bibinfo{person}{Aaron Lefohn}, {and} \bibinfo{person}{Wojciech Jarosz}.}
  \bibinfo{year}{2020}\natexlab{}.
\newblock \showarticletitle{Spatiotemporal reservoir resampling for real-time
  ray tracing with dynamic direct lighting}.
\newblock \bibinfo{journal}{\emph{ACM Transactions on Graphics (TOG)}}
  \bibinfo{volume}{39}, \bibinfo{number}{4} (\bibinfo{year}{2020}),
  \bibinfo{pages}{148--1}.
\newblock


\bibitem[\protect\citeauthoryear{Blumer, Nov{\'a}k, Habel, Nowrouzezahrai, and
  Jarosz}{Blumer et~al\mbox{.}}{2016}]%
        {blumer2016reduced}
\bibfield{author}{\bibinfo{person}{Adrian Blumer}, \bibinfo{person}{Jan
  Nov{\'a}k}, \bibinfo{person}{Ralf Habel}, \bibinfo{person}{Derek
  Nowrouzezahrai}, {and} \bibinfo{person}{Wojciech Jarosz}.}
  \bibinfo{year}{2016}\natexlab{}.
\newblock \showarticletitle{Reduced aggregate scattering operators for path
  tracing}. In \bibinfo{booktitle}{\emph{Computer Graphics Forum}},
  Vol.~\bibinfo{volume}{35}. Wiley Online Library, \bibinfo{pages}{461--473}.
\newblock


\bibitem[\protect\citeauthoryear{Golubev}{Golubev}{2018}]%
        {Golubev2018}
\bibfield{author}{\bibinfo{person}{Evgenii Golubev}.}
  \bibinfo{year}{2018}\natexlab{}.
\newblock \showarticletitle{Efficient screen-space subsurface scattering using
  Burleys normalized diffusion in real-time}. In \bibinfo{booktitle}{\emph{ACM
  SIGGRAPH Courses: Advances in Real-Time Rendering in Games Course}}.
\newblock


\bibitem[\protect\citeauthoryear{Green, Kautz, Matusik, and Durand}{Green
  et~al\mbox{.}}{2006}]%
        {green2006view}
\bibfield{author}{\bibinfo{person}{Paul Green}, \bibinfo{person}{Jan Kautz},
  \bibinfo{person}{Wojciech Matusik}, {and} \bibinfo{person}{Fr{\'e}do
  Durand}.} \bibinfo{year}{2006}\natexlab{}.
\newblock \showarticletitle{View-dependent precomputed light transport using
  nonlinear gaussian function approximations}. In
  \bibinfo{booktitle}{\emph{Proceedings of the 2006 symposium on Interactive 3D
  graphics and games}}. \bibinfo{pages}{7--14}.
\newblock


\bibitem[\protect\citeauthoryear{Halko, Martinsson, and Tropp}{Halko
  et~al\mbox{.}}{2011}]%
        {halko2011finding}
\bibfield{author}{\bibinfo{person}{Nathan Halko}, \bibinfo{person}{Per-Gunnar
  Martinsson}, {and} \bibinfo{person}{Joel~A Tropp}.}
  \bibinfo{year}{2011}\natexlab{}.
\newblock \showarticletitle{Finding structure with randomness: Probabilistic
  algorithms for constructing approximate matrix decompositions}.
\newblock \bibinfo{journal}{\emph{SIAM review}} \bibinfo{volume}{53},
  \bibinfo{number}{2} (\bibinfo{year}{2011}), \bibinfo{pages}{217--288}.
\newblock


\bibitem[\protect\citeauthoryear{Ha{\v{s}}an, Pellacini, and Bala}{Ha{\v{s}}an
  et~al\mbox{.}}{2006}]%
        {havsan2006direct}
\bibfield{author}{\bibinfo{person}{Milo{\v{s}} Ha{\v{s}}an},
  \bibinfo{person}{Fabio Pellacini}, {and} \bibinfo{person}{Kavita Bala}.}
  \bibinfo{year}{2006}\natexlab{}.
\newblock \showarticletitle{Direct-to-indirect transfer for cinematic
  relighting}.
\newblock \bibinfo{journal}{\emph{ACM transactions on graphics (TOG)}}
  \bibinfo{volume}{25}, \bibinfo{number}{3} (\bibinfo{year}{2006}),
  \bibinfo{pages}{1089--1097}.
\newblock


\bibitem[\protect\citeauthoryear{I{\c{s}}{\i}k, Mullia, Fisher, Eisenmann, and
  Gharbi}{I{\c{s}}{\i}k et~al\mbox{.}}{2021}]%
        {icsik2021interactive}
\bibfield{author}{\bibinfo{person}{Mustafa I{\c{s}}{\i}k},
  \bibinfo{person}{Krishna Mullia}, \bibinfo{person}{Matthew Fisher},
  \bibinfo{person}{Jonathan Eisenmann}, {and} \bibinfo{person}{Micha{\"e}l
  Gharbi}.} \bibinfo{year}{2021}\natexlab{}.
\newblock \showarticletitle{Interactive Monte Carlo denoising using affinity of
  neural features}.
\newblock \bibinfo{journal}{\emph{ACM Transactions on Graphics (TOG)}}
  \bibinfo{volume}{40}, \bibinfo{number}{4} (\bibinfo{year}{2021}),
  \bibinfo{pages}{1--13}.
\newblock


\bibitem[\protect\citeauthoryear{Kajiya}{Kajiya}{1986}]%
        {kajiya1986rendering}
\bibfield{author}{\bibinfo{person}{James~T Kajiya}.}
  \bibinfo{year}{1986}\natexlab{}.
\newblock \showarticletitle{The rendering equation}. In
  \bibinfo{booktitle}{\emph{Proceedings of the 13th annual conference on
  Computer graphics and interactive techniques}}. \bibinfo{pages}{143--150}.
\newblock


\bibitem[\protect\citeauthoryear{Keller}{Keller}{1997}]%
        {keller1997instant}
\bibfield{author}{\bibinfo{person}{Alexander Keller}.}
  \bibinfo{year}{1997}\natexlab{}.
\newblock \showarticletitle{Instant radiosity}. In
  \bibinfo{booktitle}{\emph{Proceedings of the 24th annual conference on
  Computer graphics and interactive techniques}}. \bibinfo{pages}{49--56}.
\newblock


\bibitem[\protect\citeauthoryear{Lagarde et~al\mbox{.}}{Lagarde
  et~al\mbox{.}}{2018}]%
        {lagarde2021unity}
\bibfield{author}{\bibinfo{person}{Sebastien Lagarde} {et~al\mbox{.}}}
  \bibinfo{year}{2018}\natexlab{}.
\newblock \bibinfo{title}{Unity High Definition Render Pipeline}.
\newblock
\newblock
\urldef\tempurl%
\url{https://unity.com/srp/High-Definition-Render-Pipeline}
\showURL{%
\tempurl}


\bibitem[\protect\citeauthoryear{Lehtinen, Zwicker, Turquin, Kontkanen, Durand,
  Sillion, and Aila}{Lehtinen et~al\mbox{.}}{2008}]%
        {lehtinen2008meshless}
\bibfield{author}{\bibinfo{person}{Jaakko Lehtinen}, \bibinfo{person}{Matthias
  Zwicker}, \bibinfo{person}{Emmanuel Turquin}, \bibinfo{person}{Janne
  Kontkanen}, \bibinfo{person}{Fr{\'e}do Durand},
  \bibinfo{person}{Fran{\c{c}}ois~X Sillion}, {and} \bibinfo{person}{Timo
  Aila}.} \bibinfo{year}{2008}\natexlab{}.
\newblock \showarticletitle{A meshless hierarchical representation for light
  transport}.
\newblock In \bibinfo{booktitle}{\emph{ACM SIGGRAPH 2008 papers}}.
  \bibinfo{pages}{1--9}.
\newblock


\bibitem[\protect\citeauthoryear{Loos, Antani, Mitchell, Nowrouzezahrai,
  Jarosz, and Sloan}{Loos et~al\mbox{.}}{2011}]%
        {loos2011modular}
\bibfield{author}{\bibinfo{person}{Bradford~J Loos}, \bibinfo{person}{Lakulish
  Antani}, \bibinfo{person}{Kenny Mitchell}, \bibinfo{person}{Derek
  Nowrouzezahrai}, \bibinfo{person}{Wojciech Jarosz}, {and}
  \bibinfo{person}{Peter-Pike Sloan}.} \bibinfo{year}{2011}\natexlab{}.
\newblock \showarticletitle{Modular radiance transfer}. In
  \bibinfo{booktitle}{\emph{Proceedings of the 2011 SIGGRAPH Asia Conference}}.
  \bibinfo{pages}{1--10}.
\newblock


\bibitem[\protect\citeauthoryear{M{\"u}ller, Rousselle, Nov{\'a}k, and
  Keller}{M{\"u}ller et~al\mbox{.}}{2021}]%
        {muller2021real}
\bibfield{author}{\bibinfo{person}{Thomas M{\"u}ller}, \bibinfo{person}{Fabrice
  Rousselle}, \bibinfo{person}{Jan Nov{\'a}k}, {and} \bibinfo{person}{Alexander
  Keller}.} \bibinfo{year}{2021}\natexlab{}.
\newblock \showarticletitle{Real-time neural radiance caching for path
  tracing}.
\newblock \bibinfo{journal}{\emph{arXiv preprint arXiv:2106.12372}}
  (\bibinfo{year}{2021}).
\newblock


\bibitem[\protect\citeauthoryear{Ng, Ramamoorthi, and Hanrahan}{Ng
  et~al\mbox{.}}{2003}]%
        {ng2003all}
\bibfield{author}{\bibinfo{person}{Ren Ng}, \bibinfo{person}{Ravi Ramamoorthi},
  {and} \bibinfo{person}{Pat Hanrahan}.} \bibinfo{year}{2003}\natexlab{}.
\newblock \showarticletitle{All-frequency shadows using non-linear wavelet
  lighting approximation}.
\newblock In \bibinfo{booktitle}{\emph{ACM SIGGRAPH 2003}}.
  \bibinfo{pages}{376--381}.
\newblock


\bibitem[\protect\citeauthoryear{Penrose}{Penrose}{1955}]%
        {penrose1955generalized}
\bibfield{author}{\bibinfo{person}{Roger Penrose}.}
  \bibinfo{year}{1955}\natexlab{}.
\newblock \showarticletitle{A generalized inverse for matrices}. In
  \bibinfo{booktitle}{\emph{Mathematical proceedings of the Cambridge
  philosophical society}}, Vol.~\bibinfo{volume}{51}. Cambridge University
  Press, \bibinfo{pages}{406--413}.
\newblock


\bibitem[\protect\citeauthoryear{Ramamoorthi and Hanrahan}{Ramamoorthi and
  Hanrahan}{2001}]%
        {ramamoorthi2001efficient}
\bibfield{author}{\bibinfo{person}{Ravi Ramamoorthi} {and} \bibinfo{person}{Pat
  Hanrahan}.} \bibinfo{year}{2001}\natexlab{}.
\newblock \showarticletitle{An efficient representation for irradiance
  environment maps}. In \bibinfo{booktitle}{\emph{Proceedings of the 28th
  annual conference on Computer graphics and interactive techniques}}.
  \bibinfo{pages}{497--500}.
\newblock


\bibitem[\protect\citeauthoryear{Ritschel, Dachsbacher, Grosch, and
  Kautz}{Ritschel et~al\mbox{.}}{2012}]%
        {ritschel2012state}
\bibfield{author}{\bibinfo{person}{Tobias Ritschel}, \bibinfo{person}{Carsten
  Dachsbacher}, \bibinfo{person}{Thorsten Grosch}, {and} \bibinfo{person}{Jan
  Kautz}.} \bibinfo{year}{2012}\natexlab{}.
\newblock \showarticletitle{The state of the art in interactive global
  illumination}. In \bibinfo{booktitle}{\emph{Computer graphics forum}},
  Vol.~\bibinfo{volume}{31}. Wiley Online Library, \bibinfo{pages}{160--188}.
\newblock


\bibitem[\protect\citeauthoryear{Saito and Takahashi}{Saito and
  Takahashi}{1990}]%
        {saito1990comprehensible}
\bibfield{author}{\bibinfo{person}{Takafumi Saito} {and}
  \bibinfo{person}{Tokiichiro Takahashi}.} \bibinfo{year}{1990}\natexlab{}.
\newblock \showarticletitle{Comprehensible rendering of 3-D shapes}. In
  \bibinfo{booktitle}{\emph{Proceedings of the 17th annual conference on
  Computer graphics and interactive techniques}}. \bibinfo{pages}{197--206}.
\newblock


\bibitem[\protect\citeauthoryear{Seyb, Sloan, Silvennoinen, Iwanicki, and
  Jarosz}{Seyb et~al\mbox{.}}{2020}]%
        {seyb20uberbake}
\bibfield{author}{\bibinfo{person}{Dario Seyb}, \bibinfo{person}{Peter-Pike
  Sloan}, \bibinfo{person}{Ari Silvennoinen}, \bibinfo{person}{Michał
  Iwanicki}, {and} \bibinfo{person}{Wojciech Jarosz}.}
  \bibinfo{year}{2020}\natexlab{}.
\newblock \showarticletitle{The design and evolution of the {{UberBake}} light
  baking system}.
\newblock \bibinfo{journal}{\emph{ACM Transactions on Graphics (Proceedings of
  SIGGRAPH)}} \bibinfo{volume}{39}, \bibinfo{number}{4} (\bibinfo{date}{July}
  \bibinfo{year}{2020}).
\newblock
\urldef\tempurl%
\url{https://doi.org/10/gg8xc9}
\showDOI{\tempurl}


\bibitem[\protect\citeauthoryear{Silvennoinen and Sloan}{Silvennoinen and
  Sloan}{2019}]%
        {Silvennoinen2019}
\bibfield{author}{\bibinfo{person}{Ari Silvennoinen} {and}
  \bibinfo{person}{Peter-Pike Sloan}.} \bibinfo{year}{2019}\natexlab{}.
\newblock \showarticletitle{Ray Guiding for Production Lightmap Baking}. In
  \bibinfo{booktitle}{\emph{SIGGRAPH Asia 2019 Technical Briefs}} (Brisbane,
  QLD, Australia) \emph{(\bibinfo{series}{SA '19})}.
  \bibinfo{publisher}{Association for Computing Machinery},
  \bibinfo{address}{New York, NY, USA}, \bibinfo{pages}{91–94}.
\newblock
\showISBNx{9781450369459}
\urldef\tempurl%
\url{https://doi.org/10.1145/3355088.3365167}
\showDOI{\tempurl}


\bibitem[\protect\citeauthoryear{Sloan, Hall, Hart, and Snyder}{Sloan
  et~al\mbox{.}}{2003}]%
        {sloan2003clustered}
\bibfield{author}{\bibinfo{person}{Peter-Pike Sloan}, \bibinfo{person}{Jesse
  Hall}, \bibinfo{person}{John Hart}, {and} \bibinfo{person}{John Snyder}.}
  \bibinfo{year}{2003}\natexlab{}.
\newblock \showarticletitle{Clustered principal components for precomputed
  radiance transfer}.
\newblock \bibinfo{journal}{\emph{ACM Transactions on Graphics (TOG)}}
  \bibinfo{volume}{22}, \bibinfo{number}{3} (\bibinfo{year}{2003}),
  \bibinfo{pages}{382--391}.
\newblock


\bibitem[\protect\citeauthoryear{Sloan, Kautz, and Snyder}{Sloan
  et~al\mbox{.}}{2002}]%
        {sloan2002precomputed}
\bibfield{author}{\bibinfo{person}{Peter-Pike Sloan}, \bibinfo{person}{Jan
  Kautz}, {and} \bibinfo{person}{John Snyder}.}
  \bibinfo{year}{2002}\natexlab{}.
\newblock \showarticletitle{Precomputed radiance transfer for real-time
  rendering in dynamic, low-frequency lighting environments}. In
  \bibinfo{booktitle}{\emph{Proceedings of SIGGRAPH 2002}}.
  \bibinfo{pages}{527--536}.
\newblock


\bibitem[\protect\citeauthoryear{Sloan, Luna, and Snyder}{Sloan
  et~al\mbox{.}}{2005}]%
        {sloan2005local}
\bibfield{author}{\bibinfo{person}{Peter-Pike Sloan}, \bibinfo{person}{Ben
  Luna}, {and} \bibinfo{person}{John Snyder}.} \bibinfo{year}{2005}\natexlab{}.
\newblock \showarticletitle{Local, deformable precomputed radiance transfer}.
\newblock \bibinfo{journal}{\emph{ACM Transactions on Graphics (TOG)}}
  \bibinfo{volume}{24}, \bibinfo{number}{3} (\bibinfo{year}{2005}),
  \bibinfo{pages}{1216--1224}.
\newblock


\bibitem[\protect\citeauthoryear{Tafuri}{Tafuri}{2019}]%
        {Tafuri2019}
\bibfield{author}{\bibinfo{person}{Sebastian Tafuri}.}
  \bibinfo{year}{2019}\natexlab{}.
\newblock \showarticletitle{Strand-based Hair Rendering in Frostbite}. In
  \bibinfo{booktitle}{\emph{ACM SIGGRAPH Courses: Advances in Real-Time
  Rendering in Games Course}}.
\newblock


\bibitem[\protect\citeauthoryear{Tsai and Shih}{Tsai and Shih}{2006}]%
        {tsai2006all}
\bibfield{author}{\bibinfo{person}{Yu-Ting Tsai} {and}
  \bibinfo{person}{Zen-Chung Shih}.} \bibinfo{year}{2006}\natexlab{}.
\newblock \showarticletitle{All-frequency precomputed radiance transfer using
  spherical radial basis functions and clustered tensor approximation}.
\newblock \bibinfo{journal}{\emph{ACM Transactions on graphics (TOG)}}
  \bibinfo{volume}{25}, \bibinfo{number}{3} (\bibinfo{year}{2006}),
  \bibinfo{pages}{967--976}.
\newblock


\bibitem[\protect\citeauthoryear{Turk and Pentland}{Turk and Pentland}{1991}]%
        {turk1991eigenfaces}
\bibfield{author}{\bibinfo{person}{Matthew Turk} {and} \bibinfo{person}{Alex
  Pentland}.} \bibinfo{year}{1991}\natexlab{}.
\newblock \showarticletitle{Eigenfaces for recognition}.
\newblock \bibinfo{journal}{\emph{Journal of cognitive neuroscience}}
  \bibinfo{volume}{3}, \bibinfo{number}{1} (\bibinfo{year}{1991}),
  \bibinfo{pages}{71--86}.
\newblock


\bibitem[\protect\citeauthoryear{Yuksel}{Yuksel}{2015}]%
        {yuksel2015sample}
\bibfield{author}{\bibinfo{person}{Cem Yuksel}.}
  \bibinfo{year}{2015}\natexlab{}.
\newblock \showarticletitle{Sample Elimination for Generating Poisson Disk
  Sample Sets}.
\newblock \bibinfo{journal}{\emph{Computer Graphics Forum (Proceedings of
  EUROGRAPHICS 2015)}} \bibinfo{volume}{34}, \bibinfo{number}{2}
  (\bibinfo{year}{2015}), \bibinfo{pages}{25--32}.
\newblock
\showISSN{0167-7055}


\end{thebibliography}

\end{document}